\documentclass{aa}  

\usepackage{graphicx}
\usepackage{txfonts}
\usepackage{lipsum}
\usepackage{subcaption}         
\usepackage{lscape}             
\usepackage{placeins}           

\usepackage{natbib}
\bibpunct{(}{)}{;}{a}{}{,} 

\usepackage{multirow}

\usepackage[colorlinks=true, linkcolor=blue, citecolor=blue, urlcolor=blue]{hyperref}

\makeatletter
\nolinenumbers
\let\linenumbers\relax

\renewcommand*\aa@idline{ } 
\renewcommand*\AALogo{}
\renewcommand*\aa@doi{}
\yearCop={\ }
\renewcommand*\aa@copyrightname{}
\renewcommand*\aa@publishlink{}

\renewcommand*\aa@numarticle{}
\AtBeginDocument{
    \renewcommand*\aa@pageof{\thepage}
}
\makeatother

\begin{document}

   \title{Low-energy ring particle accretion as the origin of Pan's equatorial ridge}


%
%
%

   \author{M. Baj\corrauth{marco.baj@mail.polimi.it}        
        \and I. Fodde\email{iosto.fodde@polimi.it}
        \and L.\,F. Civati\email{luciafrancesca.civati@polimi.it}
        \and F. Ferrari\email{fabio1.ferrari@polimi.it}
        }

   \institute{Department of Aerospace Science and Technology, Politecnico di Milano, Via La Masa 34, Milan, Italy}

\date{ }


\abstract
{Pan is a small Saturnian satellite embedded within the planet's A ring, characterized by a distinctive polygonal equatorial ridge whose formation mechanism remains debated. Among the proposed explanations, ring particle accretion offers the framework most consistent with the observed morphology, yet previous simplified models could not reproduce the ridge's complex multi-lobed morphology without ad hoc assumptions, and the cause of its latitudinal spread remains unclear.}
{This work studies the complex low-energy dynamics around Pan to verify whether the ring particle accretion paths are consistent with the observed ridge morphology.}
{We model particle motion using a circular restricted three-body problem augmented with spherical harmonic perturbations. We analyze accretion paths through grid-search simulations of trajectories transiting through the necks at $L_1$ and $L_2$, and use backward propagation from the moon's surface to determine the original orbital characteristics of the impacting material.}
{Asymmetric accreting particle populations, with inner-ring particles located closer to Pan and contributing more significantly to the impact flux, produce impact distributions that strongly correlate with the ridge polygonal structure, especially for the Saturn-facing hemisphere. The ridge latitudinal spread is recovered only when the out-of-plane position and velocity of the accreting particles are strictly limited, consistent with accretion from a thin disk such as Saturn's rings. Furthermore, analysis of the particles' impact conditions revises the previously estimated minimum accretion duration downward by an order of magnitude, to $10^4$~yr, while accreting material is shown to originate from ring particles that once populated the Encke Gap.}
{By systematically sampling admissible low-energy trajectories near Pan, our grid-search approach resolves the complex accretion dynamics that previous simplified narrow-stream models do not capture, revealing that the ridge's main morphological features emerge naturally from the local low-energy dynamical environment, without arbitrarily imposing an orbital eccentricity on Pan or invoking its orbital inclination.}


   \maketitle

\section{Introduction}  \label{sec:intro}

\begin{figure} 
   \centering
   \includegraphics[width=\hsize]{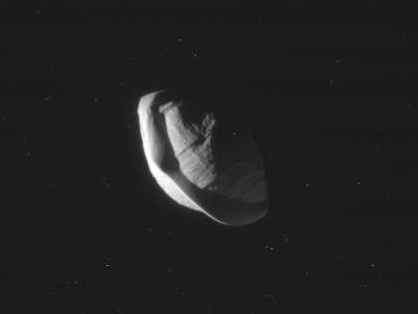}
      \caption{Image of Pan taken on March 7, 2017, by NASA's Cassini spacecraft. Image credit: NASA/JPL-Caltech/Space Science Institute.}
     \label{fig:pan_cassini_image}
   \end{figure}

The images acquired by the Cassini mission revealed that the small moons located within or near the rings of Saturn present non-trivial shapes, the origin of which is still not completely clear. One such moon is Pan, which exhibits a distinctive ridge covering its whole equator (Fig.~\ref{fig:pan_cassini_image}), earning it the nickname "ravioli-shaped moon" \citep{Leleu2018}.

The ridge appears fairly symmetric with respect to the equatorial plane, extending for about $\pm 15^\circ$ in latitude, while longitudinally five prominent lobes give the equatorial profile a characteristic polygonal outline \citep{Thomas2020}. At the resolution of the available images, the ridge appears significantly smoother and less affected by impact craters than the rest of the body, making it clearly distinct from the core and suggesting a comparatively younger origin \citep{Buratti2019, Thomas2020}. The presence of similar polygonal-shaped equatorial ridges on the nearby small moons Atlas and Daphnis \citep{Rambaux2022}, which share with Pan similar orbital and physical characteristics \citep{Ciarniello2024}, together with the absence of known comparable structures elsewhere in the Solar System, strongly suggests a common underlying mechanism for the formation of the ridges, likely related to their shared and unique dynamical environment, i.e., the close proximity to Saturn and its rings.

In this context, different ridge formation mechanisms have been proposed.
A first scenario suggests that the gravitational field generated by the rings can induce granular material freely flowing on the satellites' surfaces to accumulate along the equator \citep{Shinbrot2023}. However, this mechanism relies on the unconfirmed presence of loose surface material \citep{Buratti2019} and predicts uniform ridge structures that are inconsistent with the observed polygonal morphology. An alternative theory interprets the ridges as by-products of near head-on, low-velocity collisions between pre-existing bodies that merged to form the present-day satellites \citep{Leleu2018}. While such collisions can produce flattened shapes with equatorial ridges, they require highly constrained impact parameters and moonlets of comparable size, making it an extraordinary coincidence that all A ring moons exhibit similar ridges \citep{Shinbrot2023}. Additionally, the polygonal outline also remains unexplained, unless such complex symmetry is dismissed as an incidental result of the merging event.

A third framework proposes that the ridges formed through accretion of ring particles onto a pre-existing core \citep{Charnoz2007,Porco2007,Quillen2021}, consistent with the ridged moons' location within Saturn's rings and the coincidence of ridge and ring planes. In this scenario, the ridges' smoothness results from the small size of the accreted particles, below the resolution of Cassini's imaging system \citep{Buratti2019}, while their lower crater density reflects later deposition on a previously formed moonlet. In addition, the polygonal morphology of the ridges may stem from anisotropic accretion patterns on a tidally locked moon.

Numerical simulations confirm that ring particles can effectively reach the surface of Pan, verifying that incoming particles impact in a low-energy regime. Consequently, particles originating from orbits inner to that of Pan must pass near the $L_1$ Lagrange point before colliding with the moon, whereas those from outer orbits must approach through the vicinity of $L_2$ \citep{Charnoz2007}. \citet{Quillen2021} modeled the accumulation of ring particles on the satellite's equator,  assuming a narrow stream of particles impacting the satellite at a specific angle dictated by the particles' initial orbital radius.
However, this simplified scenario could only reproduce single-lobed or wedge-like ridge morphologies and not more complex, multi-lobed shapes, without introducing additional assumptions, such as moderate orbital eccentricity and asymmetric gaps in the ring material.
The latitudinal extent of the ridge was initially attributed to Pan's orbital inclination \citep{Charnoz2007}. However, this interpretation relied on an outdated inclination estimate approximately an order of magnitude larger than the currently accepted value \citep{Jacobson2007, Thomas2020}, and subsequent simulations demonstrated that impacts on an inclined moon  would preferentially concentrate at higher latitudes, producing U-shaped ridge morphologies inconsistent with the observed equatorial structure \citep{Quillen2021}.

Overall, while ring particle accretion offers a framework consistent with the moons' location within the ring environment, existing models struggle to reproduce the polygonal morphology of Pan's ridge or explain its latitudinal structure. Addressing these open issues and achieving a coherent and in-depth explanation of the ridge formation process would not only improve our understanding of the origin of Saturn's small moons but also help reconstruct the evolutionary history of the entire ring-moon system. Moreover, confirming the accretion hypothesis would significantly enhance the scientific relevance of Saturn's ring moons, as their ridges may represent a "fossilized" record of the surrounding ring material \citep{Charnoz2007}.

In this context, we investigated the low-energy dynamics around Pan through an in-depth exploration of the complex accretion paths arising naturally in this environment, an aspect that was only partially addressed by the simplified scenarios considered in previous studies. The aim is to assess whether the resulting spatial distribution of ring particle impacts is coherent with the observed ridge morphology, and can thus explain the polygonal longitudinal shape as well as the ridge's latitudinal extent.

By bridging the gap between simplified accretion models and the complex low-energy dynamics in the vicinity of Pan, this work provides new insights into the formation of ridged moons, with broader relevance for other low-velocity accretion processes, such as pebble accretion onto planetesimals in protoplanetary disks \citep{Ormel2017}, for which Pan can be seen as a small-scale analogue.

\section{Dynamical Framework} \label{sec:dynamical_framework}

In this section, the physical and orbital properties of Pan are introduced. Subsequently, the dynamical environment near the moon is discussed, with the relevant acceleration terms quantified and compared to identify the dominant contributions to the local acceleration field. Finally, the representative dynamical model arising from this analysis, a circular restricted three-body problem (CR3BP) with spherical harmonics perturbations, is introduced to model the accretion trajectories of ring particles.

\subsection{Physical and orbital characteristics of Pan} \label{subsec:pan_characteristics}

\begin{figure} 
   \centering
   \includegraphics[width=\hsize]{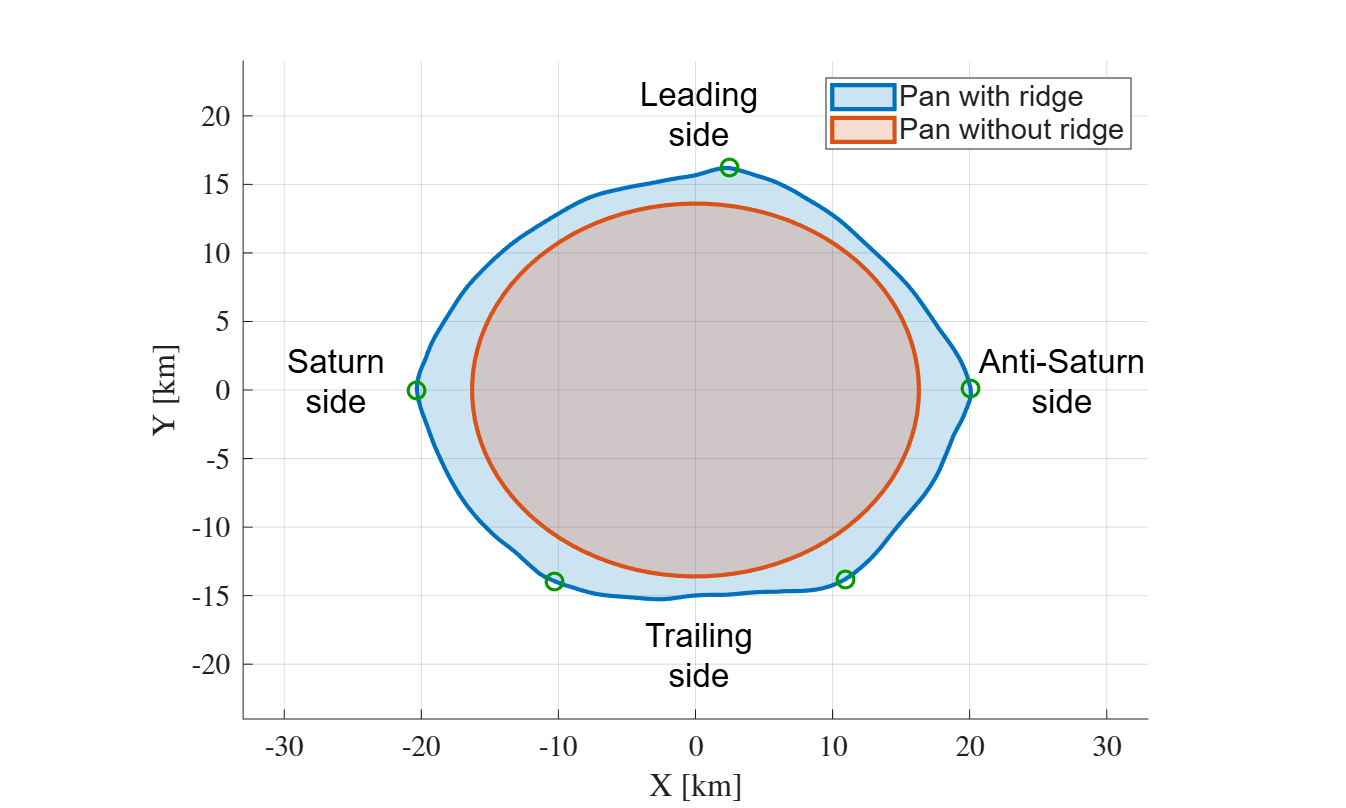}
      \caption{Comparison of the equatorial profiles of Pan with and without the equatorial ridge. The five major lobe vertices, highlighted in green, define the ridge's characteristic polygonal shape.}
         \label{fig:equat_profile}
   \end{figure}

Pan has a mean radius of 13.7~km \citep{Thomas2020} and an estimated total mass of $4.3 \times 10^{15}$~kg \citep{Weiss2009}. The images acquired by the Cassini mission enabled the reconstruction of a digital shape model of Pan \citep{Thomas2018}. Building on this, \citet{Thomas2020} characterized the shape of Pan, with the equatorial ridge removed, as a triaxial ellipsoid; we adopt this model to represent the pre-accretion shape of the moon. Figure~\ref{fig:equat_profile} displays a comparison between the equatorial cross-sections of the two models, highlighting the ridge's distinctive "polygonal" morphology, defined by five prominent lobes that give the equatorial profile a "pentagonal" appearance.

Although some studies suggest that Pan's central component may have formed from the accumulation of ring materials on a pre-existing denser core \citep{Charnoz2007,Porco2007}, in the absence of a consolidated mass distribution model, this work assumes a uniform density for Pan, consistent with the approach in \citet{Rambaux2022}. Accordingly, knowing that the ridge comprises approximately 10\% of the total volume of Pan \citep{Thomas2020}, we estimate the pre-accretion mass of Pan to be 90\% of the satellite's total mass. 
A summary of pre-accretion Pan's physical characteristics is reported in Table~\ref{tab:pan_physical_param}.

\begin{table}[ht!]
\caption{\label{tab:pan_physical_param} Summary of the pre-accretion Pan's physical characteristics.}
\centering
\begin{tabular}{lc}
\hline\hline
Parameter & Value \\
\hline
Mass, $m_\mathrm{Pan}$ (kg) & $3.87 \times 10^{15}$ \\
Longest semi-axis, $a$ (km) & 16.3 \tablefootmark{a} \\
Intermediate semi-axis, $b$ (km) & 13.6\tablefootmark{a} \\
Shortest semi-axis, $c$ (km) & 10.6\tablefootmark{a} \\
Mean radius, $R_\mathrm{Pan}$ (km) & 13.3\tablefootmark{a} \\
\hline
\end{tabular}
\tablefoot{
\tablefoottext{a}{Values from \citealt{Thomas2020}}
}
\end{table}

Dynamically, Pan is tidally locked to Saturn on a nearly circular orbit with an extremely small inclination, placing it essentially within Saturn's equatorial plane. Indeed, the moon is located in the middle of the Encke Gap, a 322 km wide empty region that the moon has cleared and maintains within the A ring, the outermost of Saturn's main rings \citep{Showalter1986,Ciarniello2024}. 
A summary of Pan's main orbital characteristics is reported in Table~\ref{tab:pan_orbital_param}.

\begin{table}[ht!]
\caption{\label{tab:pan_orbital_param} Summary of Pan's orbital characteristics.}
\centering
\begin{tabular}{lc}
\hline\hline
Parameter & Value \\
\hline
Semi-major axis, $a_\mathrm{Pan}$ (km) & 133584\\
Eccentricity & 0.0000144 \\
Inclination (deg) & 0.0001 \\
Orbital period (hours) & 13.8012  \\
\hline
\end{tabular}
\tablefoot{All values from \citealt{Jacobson2007}}
\end{table}

The past evolution of Pan's orbital elements is still largely unknown. The very short eccentricity damping time resulting from the interaction between Pan and the surrounding disk suggests that Pan's eccentricity remained very small also in the past, although temporary excitation due to orbital resonance with another satellite remains possible \citep{Quillen2021}. Pan is currently considered constrained within the Encke Gap, but it is likely that the moon experienced rapid radial migration right after its formation due to the gravitational torques generated by the nearby rings \citep{Ciarniello2024}. 
In light of these uncertainties, and considering that the exact timeframe of the accretion process is currently unknown, this work assumes that the equatorial ridge formed with Pan in its current orbital state.

\subsection{Dynamical environment around Pan} \label{subsec:perturb_analysis}

\begin{figure*} 
   \centering
   \includegraphics[width=17cm]{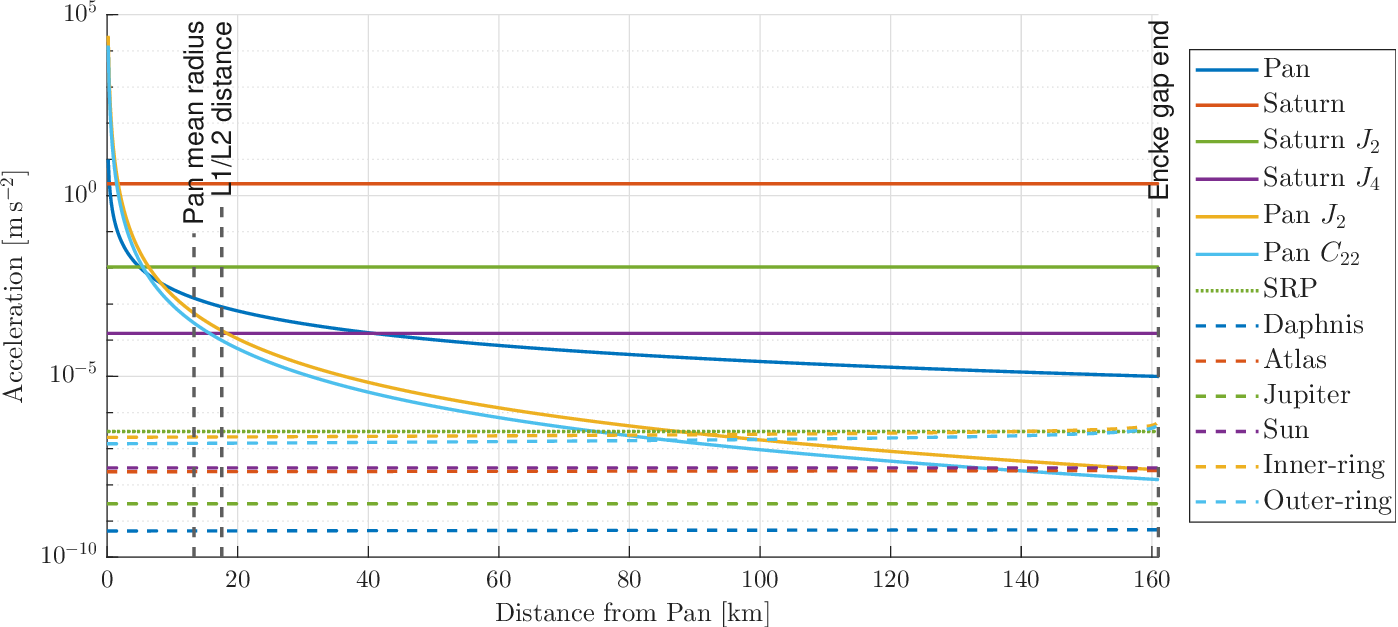}
      \caption{Comparison of the different acceleration contributions as a function of distance from Pan.}
         \label{fig:accelerations_comparison}
   \end{figure*}

To characterize the dynamical environment experienced by the accreting particles, the magnitudes of the relevant accelerations acting in the vicinity of Pan are estimated as a function of the distance from the moon, as shown in Fig.~\ref{fig:accelerations_comparison}.

The contributions considered include the gravitational acceleration of the main attractors, Saturn and Pan, as well as the perturbations introduced by the two bodies' spherical harmonics. Saturn has the shape of an oblate spheroid and, accordingly, its gravitational potential is well characterized by including its zonal harmonics of second and fourth degree, quantified respectively by the coefficients $J_2^{\mathrm{Sat}} =  16290.6\times 10^{-6}$ and $J_4^{\mathrm{Sat}} = -935.1\times 10^{-6}$, defined with respect to the reference radius $R_{\mathrm{Sat}} = 60330$~km \citep{Jacobson2022}.

As Pan is approximated as a uniform-density ellipsoid, the only non-zero coefficients in the spherical harmonic expansion of its gravitational potential up to the third degree are $J_2^\mathrm{Pan}$ and $C_{22}^\mathrm{Pan}$, given by \citep{Scheeres2012}
\begin{equation} \label{eq:pan_spherical_harmonics}
    J_2^{\mathrm{Pan}} = \frac{a^2 + b^2 - 2 c^2}{10\,R_{\mathrm{Pan}}^2}\, , \;     C_{22}^{\mathrm{Pan}} = \frac{a^2 - b^2}{20\,R_{\mathrm{Pan}}^2}\,,
\end{equation}
yielding $J_2^{\mathrm{Pan}} = 0.12772$ and $C_{22}^{\mathrm{Pan}} = 0.02282$.

The gravitational effects of Jupiter and the nearby moons Atlas and Daphnis are quantified in their maximizing configurations (i.e., at their closest approaches). The influence of Saturn's other, more distant -- albeit more massive -- moons is omitted from Fig.~\ref{fig:accelerations_comparison} for conciseness, as it has been verified to be non-dominant.
We estimate the attraction of the rings by modeling them as planar annuli with a uniform surface density of 350~kg\,m$^{-2}$, the average surface density of the A ring \citep{Miller2024}.
Since the morphology of the gap carved by Pan during the accretion epoch is uncertain, the influence of the inner and outer rings on the dynamical environment is evaluated considering the present-day architecture of the A ring. However, the order of magnitude of the resulting acceleration is verified not to vary significantly near Pan, even if the width of the Encke Gap is reduced to one quarter of its current size.

The influence of the Sun is accounted for through both its third-body gravitational perturbation on the Saturn-Pan system and the solar radiation pressure (SRP) acting on the accreting material. To quantify the maximum possible SRP effect, the acceleration is computed for a perfectly reflecting spherical particle of 1~mm in diameter, which corresponds to the smallest grain size in the A ring \citep{Miller2024}.

Comparing these contributions shows that, in the vicinity of the moon, the dynamics are dominated by the gravitational attraction of Saturn and Pan, alongside non-negligible contributions arising from their non-spherical shapes. Moreover, Fig.~\ref{fig:accelerations_comparison} indicates that throughout the Encke Gap region, when a particle is sufficiently far from the satellite, its dynamics can be approximated with reasonable accuracy by a perturbed restricted two-body model comprising Saturn and its harmonics.

\subsection{CR3BP including spherical harmonic perturbations} \label{subsec:CR3BP_perturbed} 

According to the analysis in Sect.~\ref{subsec:perturb_analysis}, the gravitational attraction of Saturn and Pan, alongside perturbations from their non-spherical gravitational fields, dominates the accretion dynamics. Additionally, the mass of an accreting ring particle is negligible compared to that of the two primary bodies, and Pan's orbit around Saturn is practically circular. Consequently, we introduce a CR3BP augmented with spherical harmonic perturbations to model the dynamics of accreting particles.

The dynamical problem is formulated in the standard synodic rotating reference frame \citep{Ross2022}, centered in the Saturn-Pan system barycenter and oriented such that both attractors remain fixed on the $x$-axis, with the secondary body (Pan) on the positive side. Additionally, the system is expressed in non-dimensional form, adopting $a_{\mathrm{Pan}}$ as the length unit $\mathrm{LU}$ and the inverse of the Keplerian mean motion as the time unit $\mathrm{TU} = \sqrt{a_{\mathrm{Pan}}^3 / G(m_{\mathrm{Pan}}+m_{\mathrm{Sat}})}$.

To account for the non-spherical gravity fields, the gravitational potentials of the primary bodies are expanded in spherical harmonics \citep{Vallado2013}, retaining the dominant higher-order terms.
For Saturn, the resulting augmented potential, expressed in the dimensionless synodic frame, is
\begin{equation} \label{eq:saturn_potential}
\begin{split}
V_{\mathrm{Sat}} = & \dfrac{1-\mu}{r_1} \Bigg[ 1 - \dfrac{1}{2} J_2^{\mathrm{Sat}} \left(\dfrac{\tilde{R}_{\mathrm{Sat}}}{r_1}\right)^2 \left( 3 \left(\dfrac{z}{r_1}\right)^2 - 1 \right) \\
& - \dfrac{1}{8} J_4^{\mathrm{Sat}} \left(\dfrac{\tilde{R}_\mathrm{Sat}}{r_1}\right)^4 \left( 35 \left(\dfrac{z}{r_1}\right)^4 - 30 \left(\dfrac{z}{r_1}\right)^2 + 3 \right) \Bigg]\,,
\end{split}
\end{equation}
where $\mu = m_{\mathrm{Pan}} / \left(m_\mathrm{Sat} + m_{\mathrm{Pan}} \right) = 6.81 \times 10^{-12}$ is the system mass ratio, while $r_1$ is the distance from Saturn and $\tilde{R}_\mathrm{Sat}$ is the reference radius used to normalize the spherical harmonic coefficients, expressed in dimensionless units. 

The gravitational potential of Pan is derived assuming that during the accretion the moon was tidally locked to Saturn, as it is at present. This assumption is supported by the observed ridge  morphology; indeed, if accretion occurred on a non-tidally locked moon, the resulting ridge would have presented a uniform altitude distribution rather than a polygonal one \citep{Quillen2021}. Under this hypothesis, the augmented potential of Pan, expressed in the dimensionless synodic frame, is given by
\begin{equation} \label{eq:pan_potential}
\begin{split}
V_{\mathrm{Pan}} = & \dfrac{\mu}{r_2} \Bigg[ 1 - \dfrac{1}{2}\, J_2^{\mathrm{Pan}} \left(\dfrac{\tilde{R}_{\mathrm{Pan}}}{r_2}\right)^2 \left( 3 \left( \dfrac{z}{r_2} \right)^2 - 1 \right) \\
& + 3 \left(\dfrac{\tilde{R}_{\mathrm{Pan}}}{r_2}\right)^2 C_{22}^{\mathrm{Pan}} \, \dfrac{ \left( x - (1-\mu) \right)^2 - y^2 }{r_2^2} \Bigg]\,,
\end{split}
\end{equation}
where $r_2$ is the distance of the particle from Pan, while $\tilde{R}_\mathrm{Pan}$ is the reference radius used to normalize the spherical harmonic coefficients, expressed in dimensionless units. 

Because both gravitational potentials are time-independent, by extending the work of \citet{Bury2020}, it can be shown that the equations of motion in the dimensionless synodic frame of the CR3BP including spherical harmonic perturbations are
\begin{equation} \label{eq:equations_of_motion}
    \begin{cases}
           \ddot x = + 2\,n\,\dot y + n^2\,x +\dfrac{\partial V_\mathrm{Sat}}{\partial x} + \dfrac{\partial V_\mathrm{Pan}}{\partial x}
 \\[3mm]
\ddot y =   - 2\, n \, \dot x + n^2\, y + \dfrac{\partial V_\mathrm{Sat}}{\partial y} + \dfrac{\partial V_\mathrm{Pan}}{\partial y} 
 \\[3mm]
    \ddot z = \dfrac{\partial V_\mathrm{Sat}}{\partial z} + \dfrac{\partial V_\mathrm{Pan}}{\partial z}
    \end{cases}\,,
\end{equation}
where $n = 1.0025253$ is the normalized mean motion of the Saturn-Pan system for the considered harmonics, as derived in Appendix~\ref{app:mean_motion}.

This system of equations possesses an integral of motion, the Jacobi constant $C$, expressed as
\begin{equation} \label{eq:jacobi_const}
    C  = n^2\left(x^2 + y^2 \right) + 2V_\mathrm{Sat} + 2V_\mathrm{Pan} - \left( \dot x^2 + \dot y^2 + \dot z^2\right)\,.
\end{equation}

This constant is negatively proportional to the energy of the third body in the rotating frame, and can therefore be interpreted as a measure of the trajectory's energy level. For a given value of $C$, Eq.~(\ref{eq:jacobi_const}) defines, for $v = \sqrt{\dot x^2 + \dot y^2 + \dot z^2} = 0$, a surface, known as the zero-velocity surface (ZVS), delimiting the region of possible motion for the third body \citep{Ross2022}.

As the particles approach the surface of Pan, the contribution of its non-spherical gravity field becomes increasingly significant. As detailed in Sect.~\ref{subsec:perturb_analysis}, we evaluated the spherical harmonic coefficients of Pan under the simplifying assumption of uniform density. However, if Pan possesses a denser inner core, as suggested by \citet{Porco2007}, the computed coefficients likely overestimate the actual gravitational perturbations. For this reason, we implement two dynamical models: the first, the "Pan-perturbed" model, includes Pan's gravitational harmonic perturbations; the second, the "Pan-unperturbed" model, neglects them and treats Pan as a point mass. These models provide upper and lower bounds on the influence of Pan's spherical harmonics, with the former likely overestimating the effects of the moon's shape on the dynamics and the latter underestimating them.

Since the two models produce different gravitational potentials for Pan, the corresponding values of Jacobi constant are not directly comparable.
For this reason, with a similar approach to that in \citet{Bury2023}, we express the values of $C$ between those of the collinear Lagrange points $L_3$ and $L_1$ (the low-energy regime) in terms of the velocity that a particle at that energy level would have while passing through the $L_1$ point. We define this "$L_1$ excess velocity" $v_{\mathrm{excess}}$ as
\begin{equation} \label{eq:excess_velocity}
    v_{\mathrm{excess}} = \sqrt{C_{L_1} - C}\,,
\end{equation}
where $C_{L_1}$ is the Jacobi constant of the $L_1$ point within the considered dynamical model.

\begin{figure} 
   \centering
   \includegraphics[width=0.9\hsize]{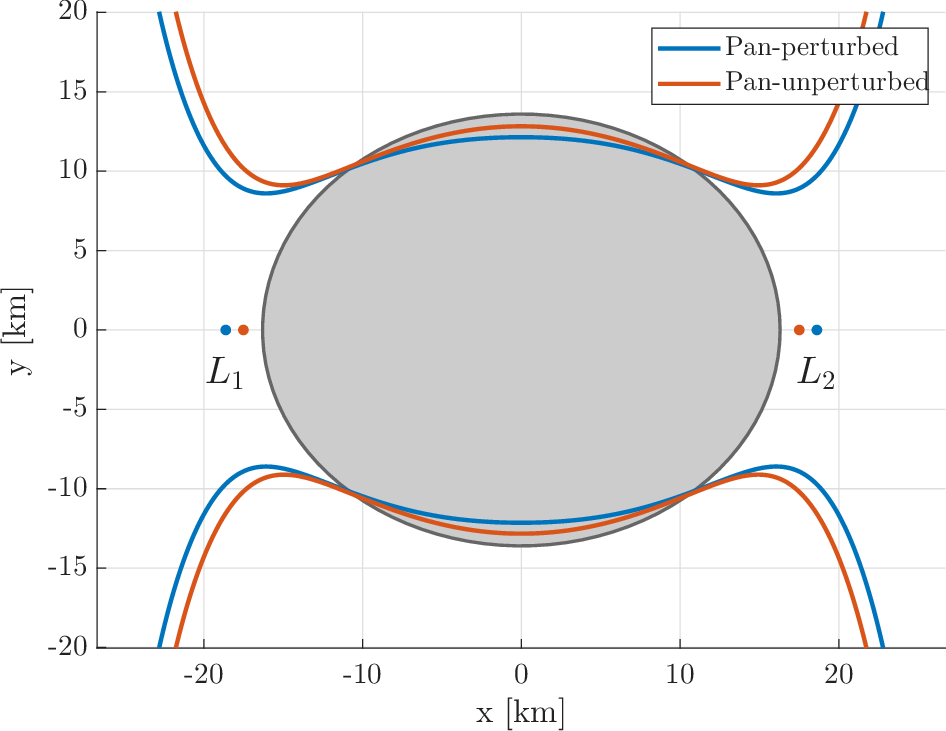}
      \caption{Lagrange points $L_1$, $L_2$ and cross-section of the ZVS obtained for $v_\mathrm{excess} = 2.00$ $\mathrm{m\,s^{-1}}$ in the Pan-perturbed model (blue) and in the Pan-unperturbed model (red).}
         \label{fig:zvs_comparison}
   \end{figure}

Table~\ref{tab:lagrange_points} summarizes the characteristics of the collinear Lagrange points for both dynamical models, while Fig.~\ref{fig:zvs_comparison} visualizes their positions and the neck opening for $v_{\mathrm{excess}} = 2.00\,\mathrm{m\,s^{-1}}$.

\begin{table*}[ht!]
\caption{\label{tab:lagrange_points} Properties of the collinear Lagrange points of the Saturn-Pan system for both the Pan-perturbed and Pan-unperturbed models.}
\centering
\begin{tabular}{clccc}
\hline\hline
Point & Model & Distance from Pan centre (km) & $C$ & $v_{\mathrm{excess}}$ ($\mathrm{m\,s^{-1}}$) \\
\hline
\multirow{2}{*}{$L_1$}  & Pan-perturbed   & 18.616 & 3.0084091021986206 & 0 \\
                        & Pan-unperturbed & 17.506 & 3.0084090949959936 & 0 \\
\hline
\multirow{2}{*}{$L_2$}  & Pan-perturbed   & 18.618 & 3.0084091021876080 & 0.0559 \\
                        & Pan-unperturbed & 17.507 & 3.0084090949868356 & 0.0510 \\
\hline
\multirow{2}{*}{$L_3$}  & Pan-perturbed   & 267168.000 & 3.0084089391431510 & 6.8044 \\
                        & Pan-unperturbed & 267168.000 & 3.0084089391431510 & 6.6524 \\
\hline
\end{tabular}
\end{table*}

\section{Numerical simulations} \label{sec:numerical_simulations}

The numerical propagation of particle trajectories from their initial state in the rings to an impact with Pan requires long integration times, making it computationally prohibitive to simulate the large ensemble of trajectories needed to statistically characterize the ring particles' impact distribution. However, ring material reaching Pan originates from adjacent, nearly circular orbits and therefore encounters the moon at low relative velocities \citep{Charnoz2007}, consistent with smooth ridge deposition rather than cratering. Thus, ring particles accreting on Pan belong to the low-energy regime, meaning they can reach the moon's surface only by transiting through the necks in the ZVS near $L_1$ and $L_2$ \citep{Ross2022}.
Consequently, we analyzed the accretion process by performing grid-search propagations of the admissible particle states at the $L_1$ and $L_2$ sections. This approach substantially reduces the integration time, as trajectories are propagated only from the neck region toward impact. We then statistically analyzed the obtained impact conditions to assess their compatibility with the ridge morphology.

To facilitate this approach, all trajectories were simulated individually, ignoring mutual interactions between the accreting particles. This assumption is justified by the low mass of the ring particles, which leads to negligible mutual gravitational effects relative to the dominant accelerations from Saturn and Pan, and by the low likelihood of particle-particle collisions over the short interaction times near the moon.

Similar approaches have been successfully implemented in the standard CR3BP to characterize the low-energy impact distributions on Iapetus \citep{Leiva2013}, as well as in perturbed CR3BP frameworks to study ballistic landing trajectories \citep{Ferrari2018,Bury2019}.

We first implemented the numerical framework in a planar configuration (Sect.~\ref{subsec:2D_grid_search}) and then extended it to three dimensions (Sect.~\ref{subsec:3D_grid_search}), using the results of the 2D case to aid the analysis. In addition, we performed planar backward propagations from the surface of Pan (Sect.~\ref{subsec:back_propagation}) to characterize the initial orbits of the impacting particles and verify their consistency with ring particle dynamical properties, thereby validating the physical relevance of the forward simulations.

\subsection{2D grid search} \label{subsec:2D_grid_search}

To reduce the computational cost of the simulations, we initially constrained the possible motion of the accreting particles to the synodic frame $xy$-plane. This assumption represents a reasonable simplification, as both Pan's ridge and the rings are predominantly confined to this plane. Moreover, Eq.~(\ref{eq:equations_of_motion}) ensures that if the initial out-of-plane components of position and velocity are zero, they remain identically zero for all times.

For both the Pan-perturbed and Pan-unperturbed dynamical models, we considered 100 equally spaced values of $C$ in the low-energy regime $[C_{L_3}, C_{L_1})$. At each energy level, we defined a set of equally spaced initial positions within the ZVS at the $L_1$ and $L_2$ sections, and for each position, we derived the corresponding velocity magnitude from the value of $C$ using Eq.~(\ref{eq:jacobi_const}). The initial-state grid was completed by sampling equally spaced velocity directions within the half-plane oriented toward Pan. The grid resolution was iteratively increased until further refinements produced negligible changes in the resulting impact distribution. Details on the adopted grid size and numerical integrator are provided in Appendix~\ref{app:numerical_setup}.

The propagation of these states was terminated upon a collision with Pan or if the trajectory crossed the $L_1$ or $L_2$ sections. Beyond these boundaries, a particle leaves Pan's sphere of influence, and its motion becomes predominantly Saturn-centric, with only a negligible fraction of the escaping trajectories coincidentally returning to the satellite's vicinity. Additionally, we imposed a maximum integration time of $20\pi$~TU, as trajectories failing to escape or impact within this interval represent a negligible fraction of the ensemble, ensuring this cutoff does not bias the resulting statistics.

We then retrieved the spherical longitude, $\lambda$, of each impact. By convention, we define $\lambda = 0^\circ$ as the anti-Saturn direction, with the angle increasing positively in the direction of Pan's rotation. For each energy level, we estimated the distribution of $\lambda$ using kernel density estimation (KDE), yielding the conditional probability density functions (PDFs) given the Jacobi constant, denoted $f^{L1}_{\lambda | C}$ and $f^{L2}_{\lambda | C}$ for the $L_1$ and $L_2$ sections, respectively.

The overall impact longitude PDFs were obtained through the marginalization of the conditional distributions over the energy distribution of the particles crossing the corresponding neck \citep{Wasserman2013}, yielding 
\begin{equation} \label{eq:marginalization_integrals}
\begin{aligned}
        f^{L1}_{\lambda}(\lambda)  &= \int  f^{L1}_{\lambda | C}(\lambda | C) \cdot f^{L1}_{C}(C)\, dC\,, \\
        f^{L2}_{\lambda}(\lambda ) &=  \int f^{L2}_{\lambda | C}(\lambda | C)  \cdot f^{L2}_{C}(C) \, dC\,,
\end{aligned}
\end{equation}
where $f^{L1}_{C}$ and $f^{L2}_{C}$ represent the PDFs of the Jacobi constant for particles crossing the $L_1$ and $L_2$ sections, respectively. These PDFs were obtained assuming the accreting particles originate from uniformly populated Saturn-centered annuli: an annulus inner to Pan's orbit, responsible for the $L_1$ particle flux, and an outer annulus representative of the $L_2$ flux.

According to \citet{Quillen2021}, a ring particle can collide with a circularly orbiting moon at the first close encounter only when its initial orbital radius differs from that of the moon by an amount $\Delta a$ satisfying \( 1.7 R_{\mathrm{H}} \lesssim | \Delta a | \lesssim 2.5 R_{\mathrm{H}}  \), where $R_{\mathrm{H}} = 19.09$~km for Pan. 
Correspondingly, we selected the boundaries of the inner and outer accreting annuli to fall within this range.

The Jacobi constant distributions in the two annuli were derived through uniform spatial sampling of ring particles, assumed to be initially on circular, Saturn-centered orbits. The particles' orbital velocity $V_\mathrm{circ}$ in the Saturn-centered inertial frame was therefore computed, including the influence of Saturn's \( J_2 \) and \( J_4 \) gravitational harmonics, as
\begin{equation} \label{eq:circular_orbital_speed}
    V_{\mathrm{circ}} = \sqrt{
\dfrac{G\,m_{\mathrm{Sat}}}{d_{\mathrm{Sat}}}
\left(
1 
+ \dfrac{3}{2} J_2^{\mathrm{Sat}} \dfrac{R_{\mathrm{Sat}}^2}{d_{\mathrm{Sat}}^2}
- \dfrac{15}{8} J_4^{\mathrm{Sat}} \dfrac{R_{\mathrm{Sat}}^4}{d_{\mathrm{Sat}}^4}
\right)}\,,
\end{equation} 
where $d_{\mathrm{Sat}}$ is the distance from the center of the planet. Each particle state was then transformed into the synodic frame (Appendix~\ref{app:frame_conversion}), and the corresponding Jacobi constant was evaluated using Eq.~(\ref{eq:jacobi_const}).
The frame transformation was performed assuming $t_0 = 0$. However, since $\mu$ is extremely small and thus the synodic frame is almost centered on Saturn, sensitivity analyses verified that using a random value for $t_0$ yields negligible changes in the results.

We then applied a KDE method to the resulting discrete sets of computed $C$ values for the inner and outer annuli, yielding $f^{L1}_{C}$ and $f^{L2}_{C}$. Using these PDFs, we numerically computed the integrals in Eq.~(\ref{eq:marginalization_integrals}) with the trapezoidal rule over the grid of equally spaced $C$ values for which trajectories were propagated.

The overall distribution of the impact longitude $f_{\lambda}$ was finally computed as the weighted sum of the PDFs relative to the two necks,
\begin{equation} \label{eq:necks_weighting}
    f_{\lambda}(\lambda) = w_{L1}\, f^{L1}_{\lambda}(\lambda) + (1-w_{L1})\, f^{L2}_{\lambda}(\lambda)\,,
\end{equation}
where the weight $w_{L1}$ is the fraction of impacts coming from the $L_1$ neck. The impact distribution $f_{\lambda}$ was evaluated across different radial configurations of the inner and outer annuli and for various values of $w_{L1}$ to identify the conditions most consistent with the observed longitudinal profile of the ridge altitude.

In the absence of detailed information on the distribution of ring-particle transit states at the neck sections, the present reconstruction assumes that all admissible entry states at a given energy level are equally probable. This approximation is justified by the strong sensitivity of the dynamical system to the initial conditions, together with the highly stochastic behavior of particles near the edges of the Encke Gap, which is expected to promote significant phase-space mixing. Additionally, the model provides a simplified description of the true accretion process: it neglects changes in Pan's morphology as accretion proceeds, and ignores potential post-impact scattering. Despite these simplifications, correlations between the simulated impact distribution and the observed ridge morphology would provide strong evidence for a causal link between the two phenomena.

While the model primarily maps the spatial distribution of impacts, the simulated impacting trajectories also provide other insights into the accretion mechanism. 
Maintaining the tidal lock during accretion requires that the torque exerted by the incoming material $T_{\mathrm{acc}}$ does not exceed Saturn's tidal torque $T_{\mathrm{tidal}}$. Following \citet{Quillen2021}, the value of $T_{\mathrm{tidal}}$ is estimated as
\begin{equation} \label{eq:tidal_torque}
    T_{\mathrm{tidal}} = \dfrac{3 G m_\mathrm{Sat}^2}{2 a_{\mathrm{Pan}}} \left( \dfrac{R_{\mathrm{Pan}}}{a_{\mathrm{Pan}}} \right)^5 \left( 0.038 \dfrac{G m_\mathrm{Pan}^2}{R_{\mathrm{Pan}}^4 \mu_\mathrm{shear}Q}  \right)\,,
\end{equation}
where the product of Pan's interior shear modulus and quality factor is assumed to be $\mu_\mathrm{shear}Q = 10^{11}$~Pa.

The accretion torque is estimated by assuming that the accreting mass flux $\dot M$ is composed of particles impacting, on average, at a distance equal to the moon's mean radius $R_\mathrm{Pan}$ with a transverse impact velocity $v_\theta$, yielding $T_{\mathrm{acc}} = \dot M R_{\mathrm{Pan}} v_\theta$. Introducing the accretion timescale $t_\mathrm{acc} = m_\mathrm{ridge} / \dot M$ reveals that the accretion torque is inversely proportional to the accretion time. A critical accretion timescale $t_\mathrm{acc}^*$ can therefore be defined by imposing $T_{\mathrm{acc}} = T_{\mathrm{tidal}}$, obtaining
\begin{equation} \label{eq:crtical_time}
    t_\mathrm{acc}^* = \dfrac{m_\mathrm{ridge}\, R_{\mathrm{Pan}} \, v_\theta}{T_{\mathrm{tidal}}}\,.
\end{equation}

While \citet{Quillen2021} evaluated $t_\mathrm{acc}^*$ using an estimated characteristic magnitude for $v_\theta$, we extend our statistical framework to reconstruct the distribution of the signed transverse velocities at impact. Using the mean of this signed distribution for $v_\theta$, our model accounts for both positive and negative angular momentum contributions when computing $t_\mathrm{acc}^*$, leading to a revised lower bound for the accretion process duration.

\subsection{3D grid search} \label{subsec:3D_grid_search}

We then extended the study of the accreting particle trajectories to three-dimensional space, allowing us to include the impact site latitudes in the analysis and assess the validity of the planar approximation. 
However, to manage the increased computational cost, we restricted the 3D grid search for both dynamical models to a single representative energy level $C$ at each section, selected based on the findings from the 2D simulations.

At each section, we sampled initial positions over a uniform grid in $(y,z)$, with $y$ spanning the ZVS maximum width and $z \in [-z_{\max}, z_{\max}]$. For each position, we obtained the velocity magnitude from $C$ using Eq.~(\ref{eq:jacobi_const}), discarding configurations yielding imaginary values, as they correspond to positions outside the admissible region of motion.

We parameterized the velocity direction using two angles: the azimuthal angle $\alpha$ in the $xy$-plane and the elevation angle $\beta$ with respect to that plane. Accordingly, we completed the grid by sampling equally spaced values for $\alpha$ within the half-plane of directions pointing toward Pan, and for $\beta \in [-\beta_{\max}, \beta_{\max}]$.

Upper bounds on the out-of-plane components, defined by $z_{\max}$ and $\beta_{\max}$, were introduced consistently with the hypothesis that the accretion process occurred after the rings had started settling into their current thin configuration \citep{Porco2007}. Based on preliminary tests on the impact latitudinal spread, we selected $z_{\max} = 2.5$~km and $\beta_{\max} = 5^\circ$. The actual distribution of vertical displacement and velocity orientation at the Lagrange point necks might be more complex than our uniform assumption; however, determining the true parameter distributions would require extensive trajectory simulations from the rings to Pan, which this work explicitly aims to avoid through the use of the grid search at the necks.

Details on the adopted grid size and numerical integrator are provided in Appendix~\ref{app:numerical_setup}. The propagation of the initial state followed the same termination criteria described in Sect.~\ref{subsec:2D_grid_search} for the 2D grid search.

We recorded each impact site using Pan's spherical longitude $\lambda$ and latitude $\theta$. For each neck, we reconstructed the spatial distribution of the impacts as a normalized two-dimensional histogram, yielding the discrete approximations of the impact PDFs, $\hat{f}_{\lambda,\theta}^{L1}$ and $\hat{f}_{\lambda,\theta}^{L2}$. We defined the coordinate grid as uniform in latitude and longitude; consequently, because Pan's surface is ellipsoidal, the corresponding bins do not encompass exactly equal surface areas. Nevertheless, as the analysis targets large-scale spatial trends and the ridge material predominantly accumulates near the equatorial region, this approximation has a negligible effect on the final particle distribution.

We then combined the discrete PDFs, assuming the fraction of impacts from the $L_1$ neck is $w_{L1}$, to obtain the overall discrete impact PDF,
\begin{equation} \label{eq:discrete_impact_weighting}
    \hat{f}_{\lambda,\theta}(\lambda,\theta)  = w_{L1}\,\hat{f}_{\lambda,\theta}^{L1}(\lambda,\theta) + (1-w_{L1})\,\hat{f}_{\lambda,\theta}^{L2}(\lambda,\theta)\,.
\end{equation}

Subsequently, we convolved this discrete PDF with a two-dimensional Gaussian filter to obtain a smoothed impact PDF ${f}_{\lambda,\theta}$. This filtering process has the dual purpose of producing a continuous distribution along the ridge surface and accounting, from a physical perspective, for the natural dispersion of particles after impact. Indeed, an accreting particle would not necessarily remain fixed at the exact impact point, but would rather undergo a local redistribution, approximated here with a Gaussian spatial profile. The filter dimensions were empirically scaled until a smooth, continuous profile was achieved, yielding standard deviations of approximately $3~\mathrm{km}$ longitudinally and $0.5~\mathrm{km}$ latitudinally.

We then used the impact distribution to generate the simulated shape of Pan after the accretion process. The height of the ridge resulting from accretion \(\Delta h\) was assumed to be proportional to the impact density at each location, such that
\begin{equation} \label{eq:ridge_altitude}
    \Delta h(\lambda,\theta) = \frac{h_{\max}}{\max\!\left( f_{\lambda,\theta} \right)} \, f_{\lambda,\theta}(\lambda,\theta)\,,
\end{equation}
where $h_{\max} = 4.03$~km is the observed maximum ridge height relative to the ellipsoidal core. The scaling factor ensures that the simulated ridge reproduces the maximum height of the actual one.

We obtained the simulated final shape of the moon by summing, along the radial direction, the local ridge thickness to the pre-accretion reference ellipsoid. The deposition was assumed to occur along the outward radial direction, which provides a reasonable approximation of the local surface normal in the equatorial region, where ridge accumulation is concentrated. Finally, we rendered the simulated shape of Pan in three dimensions and visually compared it with the actual shape of the moon to assess the agreement between the simulated and actual ridge structures.

\subsection{2D backward propagation} \label{subsec:back_propagation}

While the forward grid search characterizes the spatial impact distribution, it leaves the origin of the accreting particles uninvestigated. To ensure the physical consistency of the accretion hypothesis, we back-propagated low-energy impacting states on Pan to verify that they originate from the surrounding ring environment. By tracing their original orbital characteristics, we also identified the conditions under which injection onto a collision course is most likely. Due to the extended integration time spans required to propagate trajectories beyond the Lagrange point necks, we restricted this backward analysis to the planar approximation, as justified in Sect.~\ref{subsec:2D_grid_search}.

For both the Pan-perturbed and Pan-unperturbed dynamical models, we selected 100 equally spaced values of $C$ within the low-energy regime, $[C_{L_3}, C_{L_1})$. At each energy level, we defined a set of impact locations uniformly distributed in longitude along Pan's equator and evaluated the corresponding velocity magnitude using Eq.~(\ref{eq:jacobi_const}). We completed the ensemble of impacting states by generating an evenly spaced set of velocity directions pointing toward Pan's surface at each location.

We then back-propagated these states for 1000~TU, approximately 150 revolutions of Pan, allowing the trajectories to reach distances where the satellite's gravitational influence becomes negligible compared to Saturn's attraction. Trajectories intersecting Pan's equatorial surface during the backward integration were discarded, ensuring that only physically admissible accretion paths were retained. Details on the adopted grid resolution and numerical integrator are provided in Appendix~\ref{app:numerical_setup}.

To characterize the accreting particles' original orbits, we transformed the back-propagated trajectories into the Saturn-centered inertial reference frame, as detailed in Appendix~\ref{app:frame_conversion}, retaining the distance from Saturn $d_{\mathrm{Sat}}(t)$ and the inertial velocity vector $\vec{V}_\mathrm{inertial}(t)$.
We assumed $t_0 = 0$ for the frame transformation, since the extremely small value of $\mu$ ensures that arbitrary values of $t_0$ yield negligible differences in the final results.

Subsequently, we quantified the deviation of each back-propagated trajectory from a circular orbit, characteristic of Saturn's ring particles, using the parameter $\delta_{\%}(t)$, defined as
\begin{equation} \label{eq:delta_percentage}
    \delta_{\%}(t) = \dfrac{|\vec V_\mathrm{circ}(t) - \vec{V}_\mathrm{inertial}(t)|}{V_\mathrm{circ}(t)} \cdot 100\,,
\end{equation}
where $\vec{V}_\mathrm{circ}(t)$ is the circular orbit velocity vector at the corresponding location, with its magnitude given by Eq.~(\ref{eq:circular_orbital_speed}). 

In addition, to obtain a measure of the original orbit $d_{\mathrm{Sat}}$ and $\delta_{\%}$ independent of the specific integration time, we averaged the values of $d_{\mathrm{Sat}}(t)$ and $\delta_{\%}(t)$ over the chronologically earliest five revolutions of the particle around Saturn. Over these same five revolutions, we defined the effective orbital eccentricity as
\begin{equation} \label{eq:effective_eccentricity}
    e_\mathrm{eff} = \dfrac{\max(d_{\mathrm{Sat}}(t))-\min(d_{\mathrm{Sat}}(t))}{\max(d_{\mathrm{Sat}}(t))+\min(d_{\mathrm{Sat}}(t))}\,.
\end{equation}

We therefore characterized the original orbits of the impacting particles in terms of $d_{\mathrm{Sat}}$ and $e_\mathrm{eff}$ to assess their compatibility with Encke Gap particles. We preferred this set of parameters over the standard osculating Keplerian elements, as the latter lose their physical interpretation in an environment strongly perturbed by Saturn's zonal harmonics. Furthermore, we analyzed the variation of the mean and minimum values of $\delta_{\%}$ as a function of the energy level to identify $v_{\mathrm{excess}}$ ranges where only minor deviations from a purely circular orbit are sufficient to result in an impact with Pan.

\section{Results and discussion} \label{sec:result_and_discussion}

This section presents the results of the numerical simulations for both the Pan-perturbed and Pan-unperturbed models, discussing their implications for the formation of Pan's equatorial ridge. First, the outcomes of the 2D and 3D grid searches are evaluated to assess the compatibility between the particle accretion process and the observed ridge morphology. Next, the accretion torque is investigated, leading to an estimate of the minimum accretion timescale required to maintain the moon's tidally locked state. Lastly, the findings of the backward propagation are presented, highlighting the relationship between the impacting particles and the surrounding ring environment.

\subsection{Impact distribution and ridge morphology} \label{subsec:imapct_distribution}

Because the Saturn-Pan system is characterized by an extremely small value of $\mu$, the $L_1$ and $L_2$ equilibrium points are located at nearly equal distances from the moon and possess very similar energy levels (Table~\ref{tab:lagrange_points}). Consequently, for a given Jacobi constant, the trajectories propagated from the two necks reveal a marked symmetry with respect to the center of Pan. Additionally, this small mass ratio causes the ZVS necks to open very close to Pan's surface (Fig.~\ref{fig:zvs_comparison}), meaning that trajectories crossing the $L_1$ section predominantly impact the Saturn-facing hemisphere, whereas those transiting through $L_2$ preferentially collide with the anti-Saturn side.

As the observed longitudinal profile of the ridge is not symmetric about Pan's center, these preliminary results suggest that, to realistically reproduce the moon's morphology, asymmetries must be introduced into the accretion processes from the $L_1$ and $L_2$ regions.
Because the system's dynamical symmetry also produces nearly identical Jacobi constant distributions for symmetrically placed accreting particle annuli, the symmetry in the flux of particles crossing the two necks was broken by considering distinct radial displacements for the inner and outer accreting particle annuli.

To analyze the different accreting configurations, we systematically explored the parameter space by varying the radial displacement bounds of the annuli in increments of $0.1\,R_{\mathrm{H}}$. We evaluated the resulting impact distributions based on their correlation with the actual ridge height profile, visually comparing the number and location of the main peaks. Finally, we adjusted the relative contributions of the $L_1$ and $L_2$ fluxes to maximize the morphological similarity between the simulated and observed curves. Overall, for both the Pan-perturbed and Pan-unperturbed models, this procedure yielded the best agreement under the following conditions:
\begin{enumerate}
\item Inner-ring particles: Particles originating from the inner ring are distributed within the portion of the admissible radial displacement interval $[-2.5\,R_{\mathrm{H}}, -1.7\,R_{\mathrm{H}}]$ closest to Pan.
\item Outer-ring particles: Particles originating from the outer ring are either evenly distributed across the entire radial displacement interval $[1.7\,R_{\mathrm{H}}, 2.5\,R_{\mathrm{H}}]$ or concentrated toward its outer boundary.
\item Impact asymmetry: The fraction of impacts originating from the $L_1$ neck is slightly higher than that from the $L_2$ neck.
\end{enumerate}

Representative cases consistent with these conditions are defined for the Pan-perturbed and Pan-unperturbed models, using the parameters listed in Table~\ref{tab:2D_accreting_populations}.

\begin{table}[ht!]
\caption{\label{tab:2D_accreting_populations} Characteristics of the accreting particle populations adopted as representative cases.}
\centering
\begin{tabular}{lcc}
\hline\hline
Parameter & Pan-perturbed & Pan-unperturbed \\
\hline
Inner annulus, $\Delta a$ & $[-1.8, -1.7]\,R_{\mathrm{H}}$ & $[-1.9, -1.7]\,R_{\mathrm{H}}$ \\
Outer annulus, $\Delta a$ & $[+1.7, +2.5]\,R_{\mathrm{H}}$ & $[+1.7, +2.5]\,R_{\mathrm{H}}$ \\
Impact fraction, $w_{L1}$ & 55\% & 55\% \\
\hline
\end{tabular}
\tablefoot{The annuli bounds $\Delta a$ are expressed as radial offsets from Pan's semi-major axis, $a_{\mathrm{Pan}}$, in units of the Hill radius $R_{\mathrm{H}} = 19.09$\,km.}
\end{table}

\begin{figure*}
\sidecaption
  \includegraphics[width=12cm]{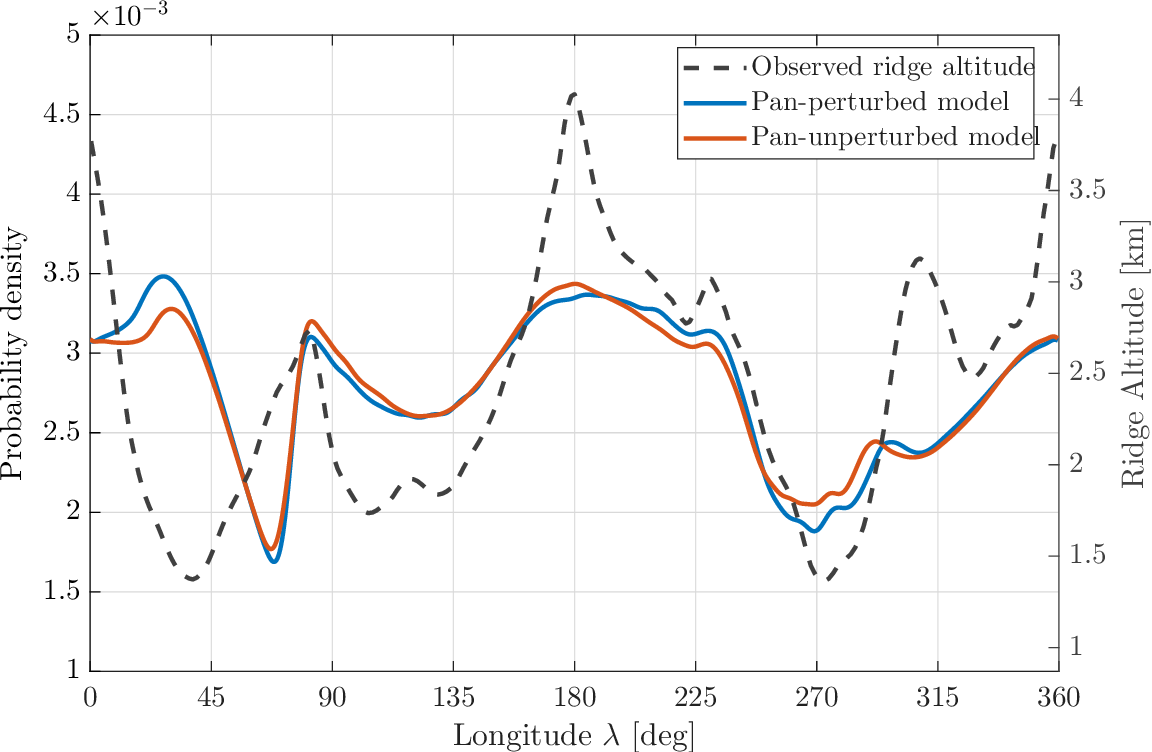}
     \caption{Overall impact longitude $\lambda$ distribution, normalized as a PDF, for the accreting particle populations defined in Table~\ref{tab:2D_accreting_populations} for the Pan-perturbed model (blue) and Pan-unperturbed models (red), plotted alongside Pan's equatorial ridge altitude as a function of longitude. For the adopted convention, $\lambda = 0^\circ$ corresponds to the anti-Saturn direction, $\lambda = 90^\circ$ to the leading side, $\lambda = 180^\circ$ to the Saturn-facing direction, and $\lambda = 270^\circ$ to the trailing side.}
     \label{fig:longitude_distribution}
\end{figure*}

Figure~\ref{fig:longitude_distribution} presents the corresponding overall impact longitudinal distributions, plotted alongside the observed equatorial ridge altitude profile. The altitude profile was obtained as the difference between the complete shape model of Pan and the reference ellipsoid approximating its central component.

A clear similarity emerges between the computed longitudinal impact distributions and the observed ridge altitude profile, with curves exhibiting the same number of main peaks and troughs. The correlation is particularly marked towards the Saturn-facing side, with the positions of the impact probability peaks at $81^\circ$, $180^\circ$, and $231^\circ$ longitude corresponding to the position of the ridge altitude maxima. Furthermore, the local minimum in impact density at $273^\circ$ coincides with the minimum ridge altitude on the trailing side. 

On the anti-Saturn hemisphere, predominantly impacted by particles transiting through the $L_2$ neck, the correlation weakens, with the positions of the peaks not exactly matching. In particular, the ridge altitude maximum at $0^\circ$ appears displaced toward the leading side in the simulated impact distribution. This discrepancy between the results obtained for the inner and outer particle accretion suggests that the particle distribution may be more complex than the one assumed for the outer ring, or that the assumption of equal probability for each initial condition at the neck may not hold for the $L_2$ section.

The Pan-perturbed and Pan-unperturbed dynamical models produce similar impact distributions. The main difference is that the Pan-unperturbed case results in slightly sharper peaks, yielding marginally better agreement with the ridge altitude profile.
Overall, the results indicate that the contribution of Pan's spherical harmonics to the main features of the longitudinal impact pattern is limited and suggest that similar results would be obtained for values of $J_2^\mathrm{Pan}$ and $C_{22}^\mathrm{Pan}$ that lie between those used for the two simulated models.

It should be remarked that the impact distribution is strictly representative of the first phases of the accretion, recalling that the model neglects the dynamical evolution of the system as material accumulates. For instance, as the portions of the ridge near the $L_1$ and $L_2$ points grow, their surfaces move closer to the neck, increasing the probability of an impact in these regions. This neglected phenomenon could explain why the ridge at $0^\circ$ and $180^\circ$ is significantly higher than the rest of the structure. Furthermore, the computed impact distribution assumes that the particles stick to the exact position where they land, while the actual process could be more complex, with the degree of scattering depending on both particle size and impact conditions.

Despite the model's limitations, the simulated impact distribution exhibits significant common features with the ridge altitude profile, demonstrating that the accretion process can produce a multi-lobed ridge structure and supporting the interpretation that the polygonal morphology of Pan's ridge is a consequence of the ring particle accretion mechanism. Additionally, our simulations indicate that the formation of a Pan-like  ridge can arise naturally from complex perturbed-CR3BP dynamics. In contrast, previous simplified dynamical scenarios could only reproduce such complex shapes by artificially introducing an orbital eccentricity, a condition difficult to justify without ongoing dynamical forcing throughout the entire accretion process \citep{Quillen2021}.

For the particle populations in Table~\ref{tab:2D_accreting_populations}, the $v_{\mathrm{excess}}$ distributions differ slightly between the two models. Outer-ring particles span approximately $4.5$-$5.8$ $\mathrm{m\,s^{-1}}$ in the Pan-perturbed case and a slightly lower range of $4.2$-$5.5$ $\mathrm{m\,s^{-1}}$ in the Pan-unperturbed case. Meanwhile, inner-ring particles cluster around $5.7$ $\mathrm{m\,s^{-1}}$ and $5.45$ $\mathrm{m\,s^{-1}}$, respectively.
Based on these values, we established the energy levels $v_{\mathrm{excess}}^{L1}$ and $v_{\mathrm{excess}}^{L2}$ for particles accreting through the $L_1$ and $L_2$ necks for the 3D grid search. Table~\ref{tab:3D_accreting_populations} summarizes these selected excess velocities along with the fraction of impacts originating from the $L_1$ neck.

\begin{table}[ht!]
\caption{\label{tab:3D_accreting_populations} Energy levels and relative impact fractions of the ring particles accreting from the $L_1$ and $L_2$ sections, selected for the 3D grid search simulations.}
\centering
\begin{tabular}{lcc}
\hline\hline
Parameter  & Pan-perturbed & Pan-unperturbed \\
\hline
$v_{\mathrm{excess}}^{L1}$ ($\mathrm{m\,s^{-1}}$) & 5.7 & 5.45 \\
$v_{\mathrm{excess}}^{L2}$ ($\mathrm{m\,s^{-1}}$) & 5.0 & 4.75 \\
$w_{L1}$ & 58\% & 58\% \\
\hline
\end{tabular}
\end{table}

\begin{figure*} 
   \centering
        \subfloat[Overall view]{
    \includegraphics[trim=30mm 30mm 30mm 30mm, clip, width=17cm]{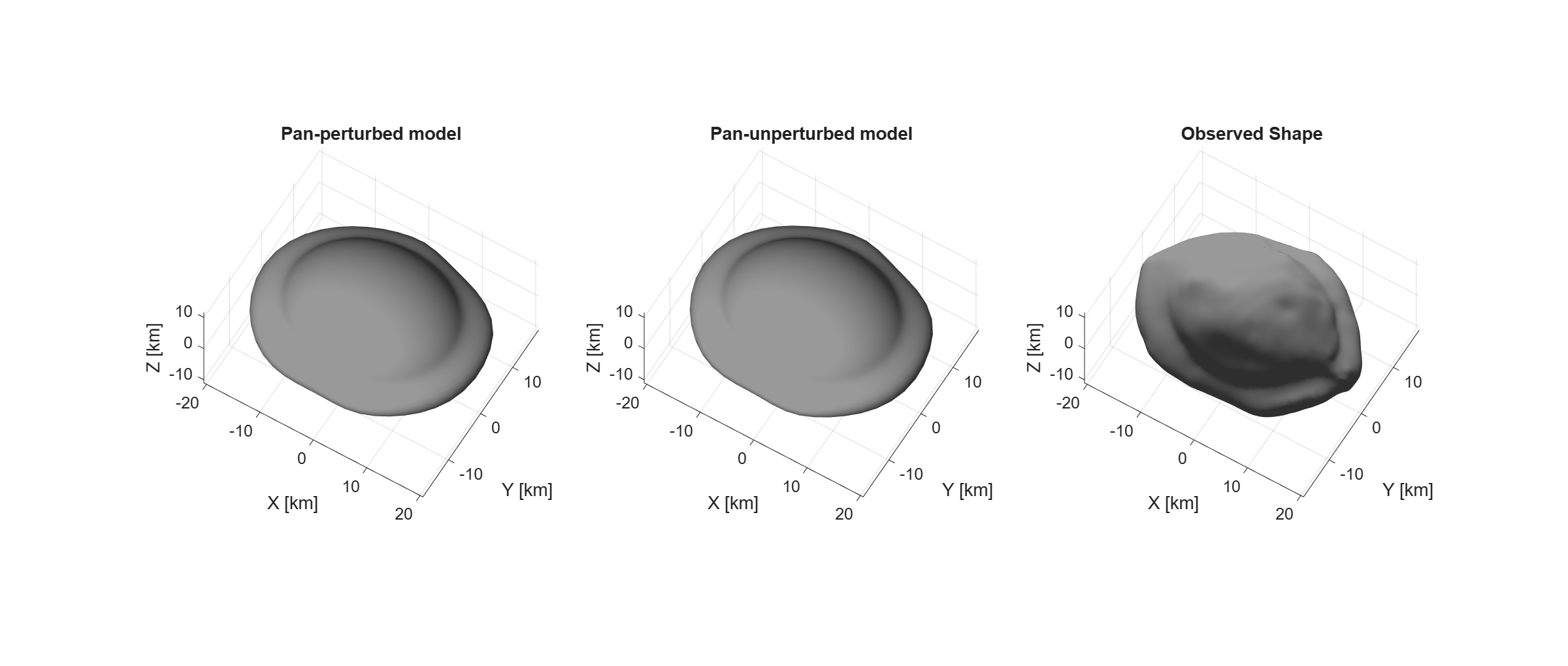}
    } \\
    \subfloat[Top view]{
        \includegraphics[width=9cm]{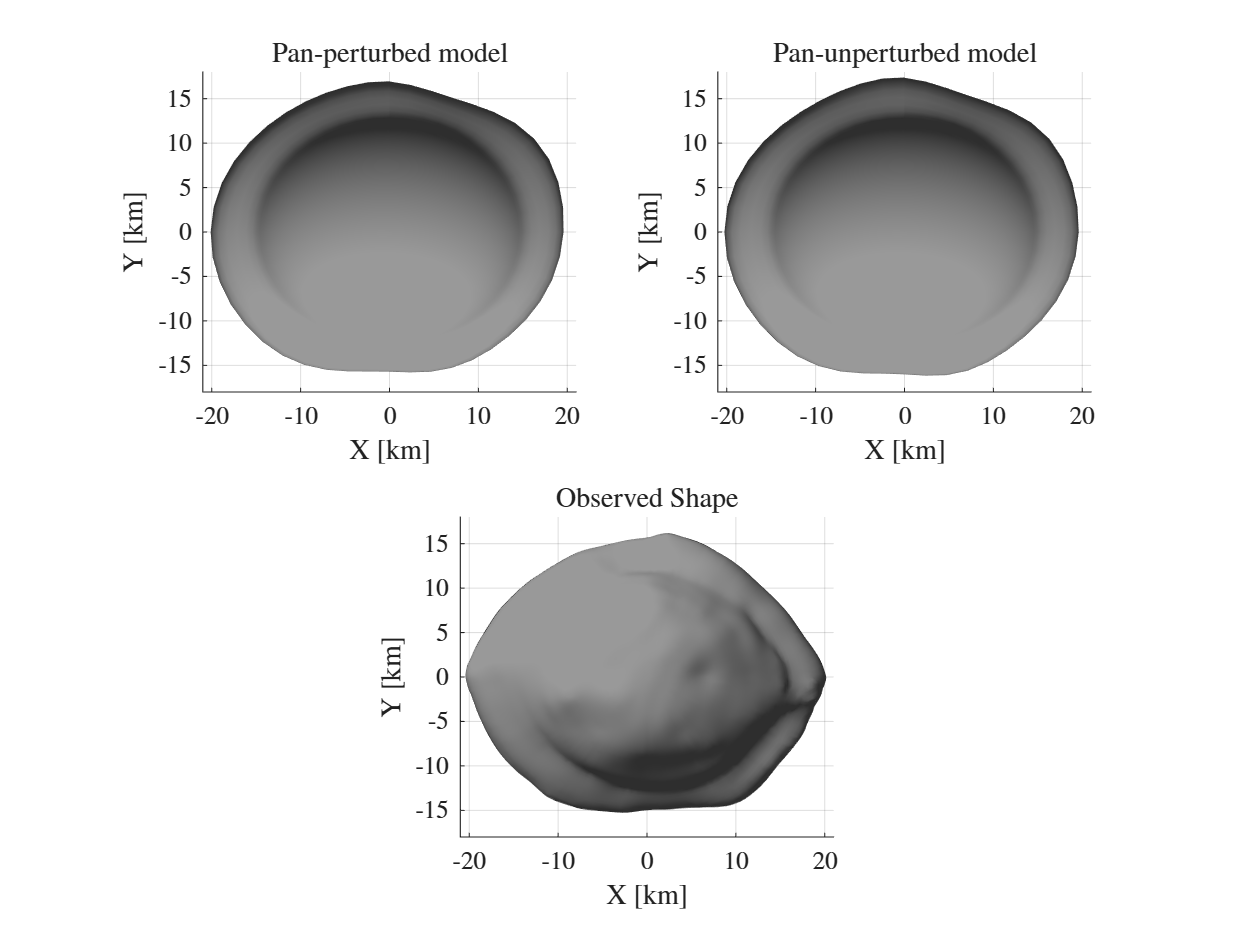}
    }
    \subfloat[Side view]{
        \includegraphics[width=9cm]{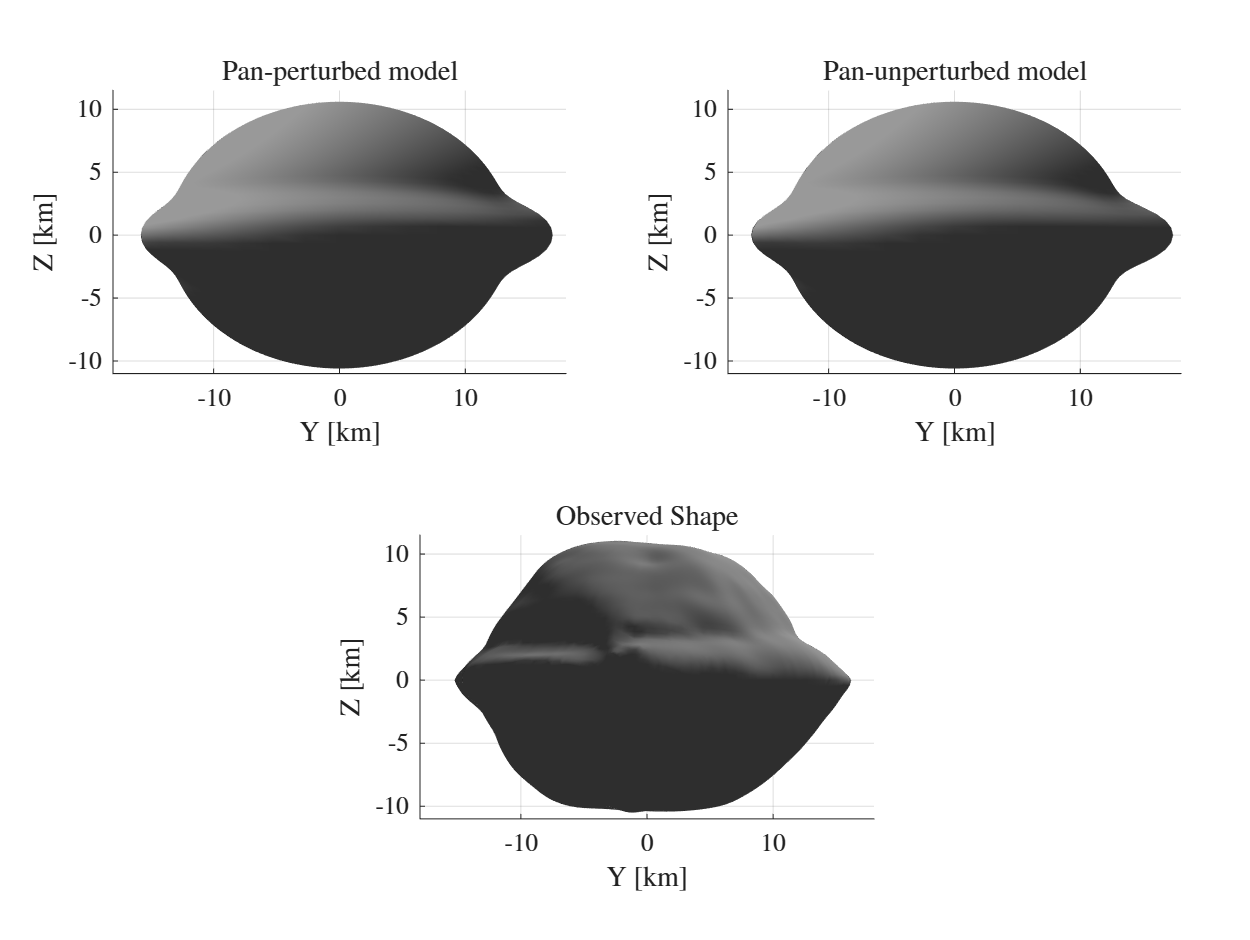}
    }
    \caption{Different views of the simulated shape of Pan resulting from the 3D grid search for the parameters reported in Table~\ref{tab:3D_accreting_populations}, alongside the observed shape of Pan for comparison.}
    \label{fig:pan_simulated_shape}
   \end{figure*}

For the selected parameters, as explained in Sect.~\ref{subsec:3D_grid_search}, the overall distribution of the impact locations resulting from the grid propagation was used to simulate the shape of Pan at the conclusion of the accretion process.
The resulting morphologies for the Pan-perturbed and Pan-unperturbed models are presented in Fig.~\ref{fig:pan_simulated_shape}, alongside the observed shape model of the moon for direct visual comparison. 

The ridge morphologies obtained with the different dynamical models are highly similar, indicating that the ring particle accretion process produces consistent outcomes regardless of whether the perturbative terms are included in Pan's gravitational potential. For both models, the simulated structures reproduce the observed shape of the moon with a high degree of visual fidelity.
In particular, the top views show very good agreement with the actual ridge morphology on the Saturn-facing and trailing sides, while a slight mismatch is observed in the position of the lobe near the anti-Saturn side, which appears shifted toward the leading edge in the simulated shape. Overall, the longitudinal distribution of the simulated altitude, and thus of the impact locations, closely aligns with the results obtained in the planar case.
This consistency validates the planar approximation and reinforces the conclusions drawn from the 2D analysis.

Concerning the side views, the simulated model captures the higher ridge altitude on the leading side relative to the trailing side, while also reproducing the overall ridge latitudinal spread with good agreement. It is important to point out that this last result stems from the prescribed bounds for the out-of-plane parameters $z_{\max}$ and $\beta_{\max}$. Moreover, the fact that the real ridge is marginally sharper than the simulated one suggests that the transit conditions at the necks may be more accurately described by a normal distribution centered at $z = 0$ and $\beta = 0$, rather than the uniform distribution assumed in our grid search.

Nevertheless, the necessity of strictly limiting the particles' out-of-plane components to reproduce the observed geometry strongly indicates that the ridge accreted from a very thin disk, such as Saturn's rings.
Additionally, our simulation results show that the latitudinal shape of the ridge can be fully accounted for by a natural dispersion of the ring particles' out-of-plane components. This removes the need to invoke the moon's orbital inclination as the primary driver of the latitudinal spread, a hypothesis that has been shown to produce U-shaped ridges where the altitude maximizes above and below the equator, contrary to the observed equator-centered bulge \citep{Quillen2021}. In this context, the higher latitudinal spread of Atlas's ridge could indicate that its formation occurred further in the past, at an earlier stage of the ring-thinning process \citep{Esposito2010}, although it is not possible to exclude a priori that Atlas's higher orbital inclination and eccentricity \citep{Jacobson2007} also played a substantial role.

In summary, the exceptional morphological match between our simulated model and Pan's observed shape confirms that ring particle accretion dynamics can account for the essential features of Pan's ridge. Most notably, this framework explains its polygonal structure, which alternative theories of merging \citep{Leleu2018} and surface granular flow reshaping \citep{Shinbrot2023} leave unexplained.

\subsection{Minimum accretion timescale} \label{subsec:min_accretion_time}

\begin{figure} 
   \centering
   \includegraphics[width=\hsize]{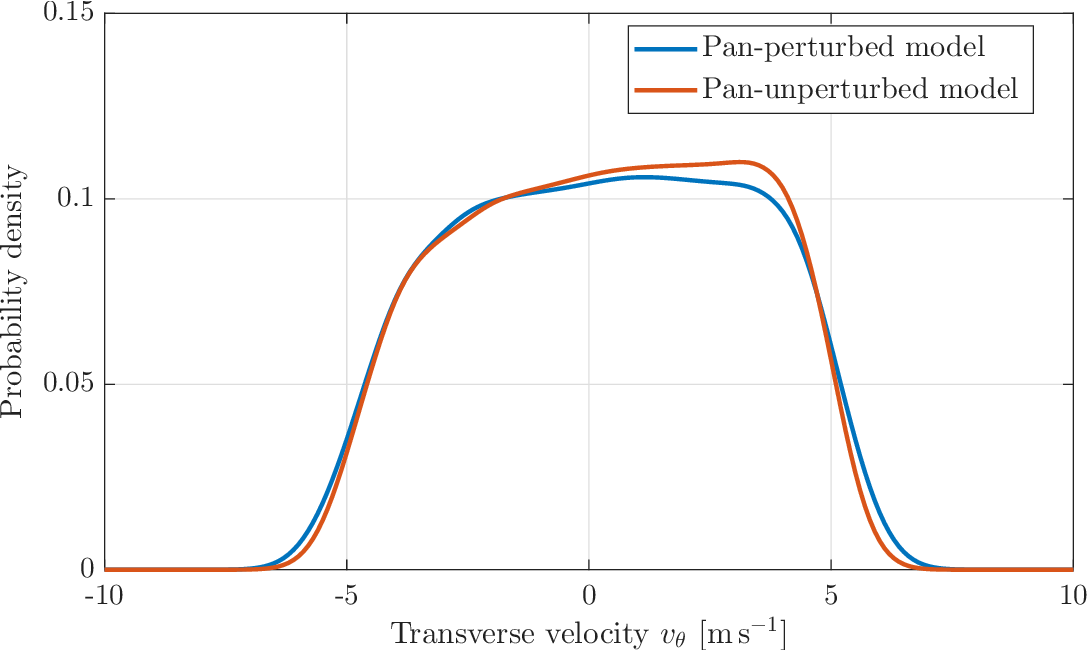}
      \caption{Overall impact transverse velocity $v_{\theta}$ distribution, normalized as a PDF, for the accreting particle populations in Table~\ref{tab:2D_accreting_populations} for the Pan-perturbed model (blue) and Pan-unperturbed model (red).}
         \label{fig:transverese_velocity_distribution}
   \end{figure}

For the same accreting particle populations described in Table~\ref{tab:2D_accreting_populations}, we also computed the corresponding impact transverse velocity $v_{\theta}$ distributions, as displayed in Fig.~\ref{fig:transverese_velocity_distribution}. For both the Pan-perturbed and the Pan-unperturbed dynamical models, the distribution is asymmetric, yielding a positive mean transverse velocity of $0.33~\mathrm{m\,s^{-1}}$ and $0.36~\mathrm{m\,s^{-1}}$, respectively.

 These non-zero mean values imply a net torque exerted on the moon by the accreting material. According to Eq.~(\ref{eq:crtical_time}), this yields a critical accretion timescale -- at which the accretion torque balances the tidal torque -- of $1.87 \times 10^3$~yr for the Pan-perturbed model and $2.04 \times 10^3$~yr for the Pan-unperturbed model. 
For comparison, \citet{Quillen2021} obtained a critical accretion timescale for Pan of $2 \times 10^4$~yr. This is consistent with the fact that their estimated transverse impact velocity of approximately $3.6\,\mathrm{m\,s^{-1}}$ is about one order of magnitude greater than the mean transverse velocities we obtained here by explicitly accounting for both positive and negative velocity contributions, as explained in Sect.~\ref{subsec:2D_grid_search}.

Assuming the accretion timescale must be at least one order of magnitude greater than the critical accretion timescale to preserve the ridge's non-axisymmetric structures, consistent with the approach in \citet{Quillen2021}, our results suggest that the minimum accretion duration required for ridge formation may be lower than previously estimated, decreasing from approximately $10^5$~yr \citep{Quillen2021} to $10^4$~yr. While this revised estimate makes the temporal constraint on accretion less stringent, the approximate $10^4$-year timescale still significantly exceeds Pan's eccentricity damping time. Analytical models estimate this damping time to be approximately 2000~yr \citep{Hahn2008}, while N-body simulations suggest the eccentricity damps in just about 100~yr \citep{Bromley2013}. Consequently, assuming Pan accreted its ridge while on an eccentric trajectory remains physically implausible without postulating a continuous external excitation mechanism, such as resonance with another moon.

\subsection{Origin of the accreting material} \label{subsec:material_origin}

\begin{figure}
   \centering
   \includegraphics[width=\hsize]{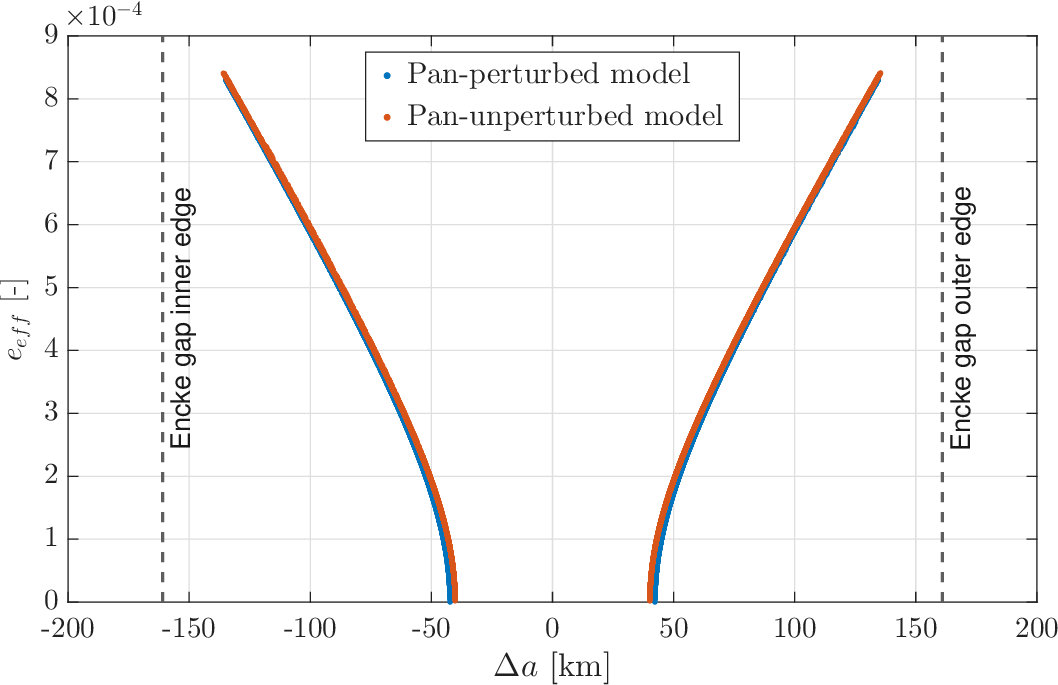}
      \caption{Relationship between the back-propagated trajectories' radial offset from Pan's orbit $\Delta a$ and their effective eccentricity $e_\mathrm{eff}$ for $v_\mathrm{excess} = 5$ $\mathrm{m\,s^{-1}}$ in both the Pan-perturbed (blue) and Pan-unperturbed (red) models.}
         \label{fig:distance_eccentricity_relation}
   \end{figure}

We characterized each back-propagated impacting trajectory by its original radial offset from Pan's orbit, $\Delta a = d_\mathrm{Sat} - a_\mathrm{Pan}$, and its effective eccentricity, $e_\mathrm{eff}$. Figure~\ref{fig:distance_eccentricity_relation} shows the combination of these two parameters for each back-propagated trajectory at $v_{\mathrm{excess}} = 5\,\mathrm{m\,s^{-1}}$, thereby mapping the initial orbital conditions in the ring that can lead to accretion at this energy level.

A very strong correlation can be noted between the radial offset and the eccentricity, with points clustering near two distinct curves associated with particles inside and outside Pan's orbit. These curves exhibit symmetric behavior with respect to Pan's orbital radius, highlighting the symmetry of trajectories passing through the two necks at the same energy level. The observed correlation can be attributed to the fixed value of the Jacobi constant, which inherently couples a particle's position with the magnitude of its velocity in the synodic frame. The effective orbital eccentricity increases with the distance from Pan's orbit, with the most circular orbits corresponding to initial distances situated immediately adjacent to the satellite's orbit. The orbital trends for the Pan-perturbed and Pan-unperturbed models are highly similar, with the only minor difference being that, when Pan's gravitational perturbations are neglected, the most circular orbits correspond to trajectories slightly closer to Pan's orbital radius.
Analogous results are obtained for all the excess velocities in the low-energy regime, with the values of effective eccentricity remaining consistently small and never exceeding $10^{-3}$.

\begin{figure} 
   \centering
   \includegraphics[width=\hsize]{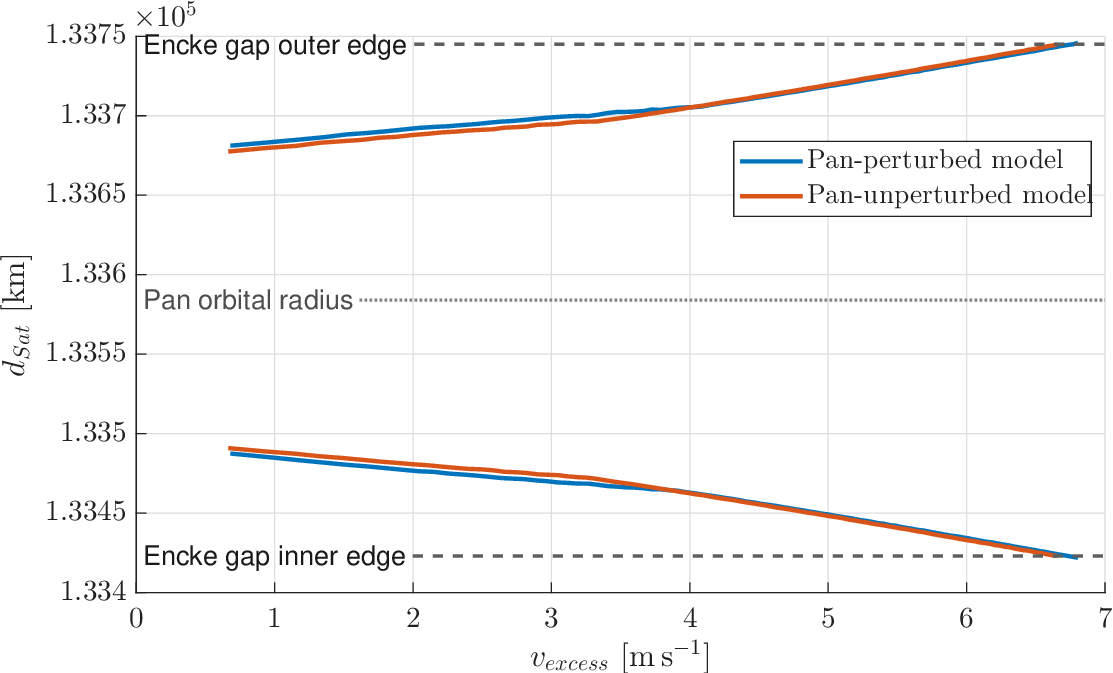}
      \caption{Initial orbital distance from Saturn $d_{\mathrm{Sat}}$ as a function of particle excess velocity $v_{\mathrm{excess}}$ for orbits corresponding to the maximum positive and negative $\Delta a$ from Pan in the Pan-perturbed (blue) and Pan-unperturbed (red) models.}
         \label{fig:max_distance}
   \end{figure}

Figure~\ref{fig:max_distance} reports, as a function of the excess velocity, the values of $d_\mathrm{Sat}$ corresponding to the inner and outer ring impacting particles that have the maximum offset from Pan's orbital radius for both the Pan-perturbed and Pan-unperturbed models. 
This plot clearly indicates that, in the low-energy regime, practically all the impacting trajectories originated from distances within the Encke Gap, consistent with the hypothesis that the ridge formed through the low-energy accretion of particles that once populated this region. Again, a strong symmetry can be noticed between the results obtained for the orbits interior and exterior to that of Pan.

\begin{figure} 
   \centering
   \includegraphics[width=\hsize]{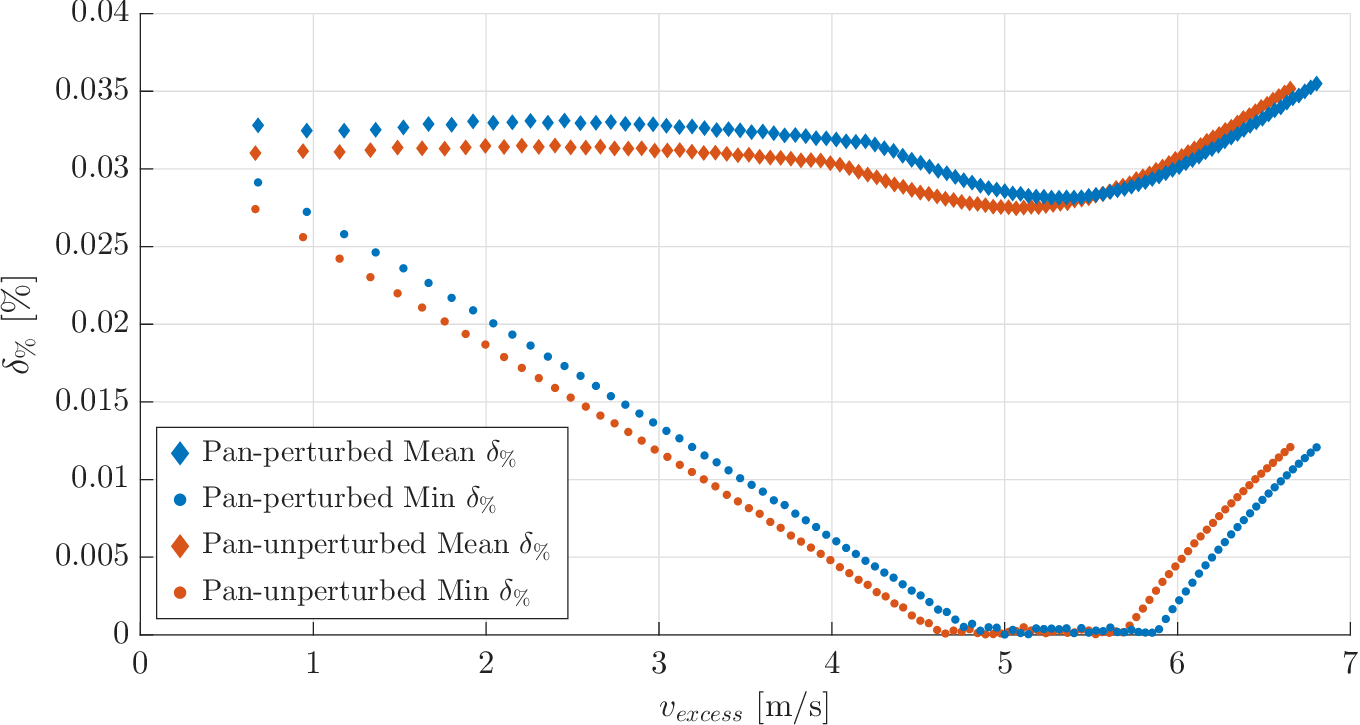}
      \caption{Minimum (dots) and mean (diamonds) values of the parameter $\delta_\%$ as a function of the excess velocity $v_{\mathrm{excess}}$ for the Pan-perturbed (blue) and Pan-unperturbed (red) models.}
         \label{fig:delta_parameters}
   \end{figure}

Finally, we quantified the deviation of the back-propagated trajectories from the circular orbits characteristic of the ring particles as a function of the energy level. We achieved this by plotting the minimum and mean values of the parameter $\delta_\%$ obtained for each considered value of $v_{\mathrm{excess}}$, as shown in Fig.~\ref{fig:delta_parameters}. The mean values of $\delta_\%$ remain consistently low, never exceeding 0.035\%, which is highly compatible with accretion scenarios involving ring particles only slightly perturbed from their original circular orbits. The Pan-perturbed and Pan-unperturbed models exhibit a similar trend, confirming that Pan's gravitational harmonics have only a minor effect on the accretion dynamics.

More interestingly, the distribution of the minimum values of $\delta_\%$ approaches zero, corresponding to originally circular particle orbits, for a distinctive region of energy levels. This occurs for $v_{\mathrm{excess}}$ values spanning approximately 4.85-5.85 $\mathrm{m\,s^{-1}}$ in the Pan-perturbed model, and 4.65-5.65  $\mathrm{m\,s^{-1}}$ in the Pan-unperturbed case. Notably, the energy levels that produced the best agreement between impact locations and ridge shape in the 3D grid search -- 5.7 and 5 $\mathrm{m\,s^{-1}}$ for the Pan-perturbed case and 5.45 and 4.75 $\mathrm{m\,s^{-1}}$ for the Pan-unperturbed case (Table~\ref{tab:3D_accreting_populations}) -- fall directly within the respective ranges where the minimum $\delta_\%$ approaches zero. 
These results not only validate the assumptions underlying the grid search but also suggest that the prevalence of impacts within this energy range is not necessarily due to a higher local particle density but rather could be the result of a lower required velocity perturbation to induce a particle to leave its initial circular orbit and collide with Pan. The velocity perturbation could then result from the mutual interaction between the ring particles or from the third-body perturbations introduced by the presence of Pan.
It is worth noting that Fig.~\ref{fig:delta_parameters} shows some noise in the region where the values of $\delta_\%$ approach zero. This is likely a consequence of the discretized nature of the grid of states used for the back-propagation; a finer sampling of these states would likely cause this set of values to converge more smoothly toward zero, reducing the observed scatter.

In summary, the characterization of these low-energy impacting trajectories demonstrates that they originate from ring particles previously populating the Encke Gap. These findings are highly consistent with the hypothesis that Pan's characteristic ridge formed through the low-energy accretion of local ring material. 
Furthermore, the agreement between the backward propagation analysis and the 3D grid search reinforces the link between the ridge morphology and Pan's interaction with the surrounding ring environment.

\section{Conclusions} \label{sec:conclusions}

This research investigated the ring particle accretion mechanism on Pan, analyzing the low-energy dynamics in the vicinity of the moon. Although the trajectories passing through the $L_1$ and $L_2$ necks exhibit strong symmetrical behavior, we show that accretion from non-symmetric particle populations accurately reproduces the moon's polygonal morphology. In particular, when the inner-ring accreting population is located closer to Pan than the outer-ring population, and impacts occur slightly more frequently from the $L_1$ side, the resulting impact distribution yields strong morphological agreement with the actual ridge altitude distribution, with only a minor deviation in the lobe position on the anti-Saturn side. Crucially, our statistical grid-search approach successfully captures the perturbed-CR3BP environment near Pan, showing that the multi-lobed ridge arises naturally from the local accretion dynamics without the artificially imposed orbital eccentricity required by earlier simplified narrow-stream accretion models \citep{Quillen2021}.

We found an impact latitudinal spread compatible with that of the ridge by limiting the particles' vertical displacement and vertical velocity component at the necks, confirming that the accretion process occurred after the rings had started settling into their current thin disk. Moreover, considering particles passing through the necks with small out-of-plane components enabled the reproduction of a ridge with the maximum altitude at the equatorial section, avoiding the complications associated with explanations relying on Pan's orbital inclination.

By extending our statistical analysis to the particle impact transverse velocities, we revised the minimum accretion timescale from the previous estimate of $10^5$ yr \citep{Quillen2021} to $10^4$ yr. 
Furthermore, we found that nearly all low-energy impacting trajectories originate from low-eccentricity orbits within the Encke Gap, indicating that the ridge accreted from the ring material that once populated this region.

Overall, the outcomes of this work demonstrate a correlation between the low-energy dynamics in the vicinity of Pan and the characteristics of its peculiar equatorial ridge, thereby reinforcing the hypothesis that this feature formed via a ring particle accretion mechanism.

Nonetheless, several avenues for future research remain open. These include refining the current model by relaxing simplifying assumptions, such as the uniform distribution of incoming particles at the necks. Future work should also investigate neglected phenomena, including Pan's possible past radial migration, the evolving ridge topography during accretion, and more realistic impact physics. These refinements could resolve our lower accuracy in reproducing the ridge shape on the anti-Saturn side and better characterize the overall accretion dynamics. Lastly, performing similar dynamical characterizations of accreting trajectories for other ridged moons, namely Atlas and Daphnis, could further strengthen confidence in the particle accretion hypothesis.

\begin{acknowledgements}
Co-funded by the European Union (ERC, TRACES, 101077758).
\end{acknowledgements}

%

\bibliographystyle{bibtex/aa} 
\bibliography{bibliography} 

\begin{appendix}

\section{Saturn-Pan system mean motion} \label{app:mean_motion}

As detailed in \citet{Bury2020}, the value of the normalized mean motion $n$ for the CR3BP including spherical harmonic perturbations can be retrieved by equating the kinematic expression of the relative acceleration $\ddot{\vec{r}}_{\mathrm{rel}}$ between the main attractors, expressed in terms of their relative position vector $\vec{r}_{\mathrm{rel}}$,
\begin{equation} \label{eq:kin_acc}
     \ddot{\vec{r}}_{\mathrm{rel}} = -n^2 \, \vec{r}_{\mathrm{rel}}\,,
\end{equation}
with the corresponding dynamical expression,
\begin{equation} \label{eq:dyn_acc}
     \ddot{\vec{r}}_{\mathrm{rel}} = \ddot{\vec{r}}_{\mathrm{Pan}} - \ddot{\vec{r}}_{\mathrm{Sat}}\,,
\end{equation}
where $\ddot{\vec{r}}_{\mathrm{Pan}}$ and $\ddot{\vec{r}}_{\mathrm{Sat}}$ denote, respectively, the gravitational accelerations experienced by Pan and Saturn due to their mutual attraction.
These accelerations can be easily retrieved, in their normalized form, computing the gradient of the potentials Eq.~(\ref{eq:saturn_potential}) and Eq.~(\ref{eq:pan_potential}), and recalling that the distance between the two bodies is normalized to 1, yielding
\begin{equation} \label{eq:pan_acc}
    \ddot{\vec{r}}_{\mathrm{Pan}}  =  - \left(1-\mu \right) \left( 1 + \dfrac{3}{2} J_2^{\mathrm{Sat}} \tilde{R}_{\mathrm{Sat}}^2 - \dfrac{15}{8} J_4^{\mathrm{Sat}} \tilde{R}_{\mathrm{Sat}}^4  \right) \vec{r}_{\mathrm{rel}}\,, 
\end{equation}
\begin{equation} \label{eq:saturn_acc}
    \ddot{\vec{r}}_{\mathrm{Sat}}  =  \mu \left( 1 + \dfrac{3}{2} J_2^{\mathrm{Pan}} \tilde{R}_{\mathrm{Pan}}^2 +9 C_{22}^{\mathrm{Pan}}\tilde{R}_{\mathrm{Pan}}^2  \right) \vec{r}_{\mathrm{rel}}\,.
\end{equation}

Substituting these expressions into Eq.~(\ref{eq:dyn_acc}) and combining them with Eq.~(\ref{eq:kin_acc}) leads to
\begin{equation} \label{eq:mean_motion}
\begin{split}
    n = &  \Bigg[ 1  + \left( 1 -\mu \right) \left( \dfrac{3}{2} J_2^{\mathrm{Sat}} \tilde{R}_{\mathrm{Sat}}^2  - \dfrac{15}{8} J_4^{\mathrm{Sat}} \tilde{R}_{\mathrm{Sat}}^4 \right) \\
     & + \mu \left( \dfrac{3}{2} J_2^{\mathrm{Pan}} \tilde{R}_{\mathrm{Pan}}^2  + 9 C_{22}^{\mathrm{Pan}}\tilde{R}_{\mathrm{Pan}}^2 \right) \Bigg]^\frac{1}{2} \,.
\end{split}
\end{equation}

Evaluating this expression with the physical parameters defined in Sect.~\ref{subsec:perturb_analysis} yields $n = 1.0025253$. For comparison, the normalized value of the measured mean motion of Pan around Saturn is equal to 1.0025209 \citep{Jacobson2007}, confirming that the implemented model accurately captures the actual planet–moon dynamics.
It can also be noted that, due to the extremely low value of $\mu$, the contribution of Pan's spherical harmonics to the numerical result is completely negligible. As expected, removing all perturbative terms recovers the standard unperturbed CR3BP value of $n = 1$.

\section{Simulations setup} \label{app:numerical_setup}

The grid dimensions selected for the 2D forward propagation are listed in Table~\ref{tab:2D_num_param}, where $n_\mathrm{pos}$ indicates the number of positions considered and $n_\mathrm{vel}$ the number of velocity orientations considered at each position.

\begin{table} [ht!]
\caption{\label{tab:2D_num_param} Summary of the numerical setup employed for the 2D grid search.}
\centering
\begin{tabular}{lc}
\hline\hline
Parameter & Value \\
\hline
 $n_\mathrm{pos}$ & $2\,500$\\
$n_\mathrm{vel}$&  100\\
Total states per $C$ value (single neck) & $250\,000$ \\
Number of $C$ values & 100 \\
Total states per dynamical model & $5 \times 10^7$ \\
\hline
\end{tabular}
\end{table}

The grid dimensions selected for the 3D forward propagation are listed in Table~\ref{tab:3D_num_param}, where $n_{\mathrm{pos},y}$ and $n_{\mathrm{pos},z}$ indicate the number of positions along the $y$- and $z$-directions, respectively, while $n_\alpha$ and $n_\beta$ represent the number of in-plane and out-of-plane velocity angles considered at each position. Owing to the symmetry of the equations of motion with respect to the $xy$-plane, only initial conditions with positive vertical displacements were numerically propagated; the trajectories for their negative counterparts were directly reconstructed via symmetry.

\begin{table}[ht!]
\caption{\label{tab:3D_num_param} Summary of the numerical setup employed for the 3D grid search.}
\centering
\begin{tabular}{lc}
\hline\hline
Parameter & Value \\
\hline
 $n_\mathrm{pos,y}$ & $500$\\
 $n_\mathrm{pos,z}$ & $20$ \\
$n_{\alpha}$&  50 \\
 $n_{\beta}$  & 50 \\
Total states per $C$ value (single neck)\tablefootmark{a} & $\sim 2.5 \times 10^7$ \\
Total states per dynamical model \tablefootmark{a} & $\sim 5 \times 10^7$ \\
\hline
\end{tabular}
\tablefoot{
\tablefoottext{a}{These are approximate values, as a small portion of the initial positions ($\lesssim$$1\%$ for the values of $C$ used) fall outside the Lagrange point neck.}
}
\end{table}

The sampling resolution selected for the backward propagation is detailed in Table~\ref{tab:backward_num_param}, where $n_\mathrm{pos}$ indicates the number of impact locations uniformly distributed along Pan's equator, and $n_\mathrm{vel}$ represents the number of velocity orientations considered at each location. Due to the significantly longer integration times required for the backward propagation compared to the forward searches, the total number of sampled states was intentionally reduced to maintain a manageable computational load.

\begin{table}[ht!]
\caption{\label{tab:backward_num_param} Summary of the numerical setup employed for the backward propagation.}
\centering
\begin{tabular}{lc}
\hline\hline
Parameter & Value \\
\hline
 $n_\mathrm{pos}$ & $100$\\
 $n_\mathrm{vel}$ & $100$\\
Total states per $C$ value & $10\,000$ \\
Number of $C$ values & 100 \\
Total states per dynamical model & $1 \times 10^6$ \\
\hline
\end{tabular}
\end{table}

All the states were then propagated using a variable-order Adams-Bashforth-Moulton integrator, implemented via the MATLAB function  \texttt{ode113} \citep{Shampine1997}. To ensure high numerical accuracy, relative and absolute tolerances were respectively set to $10^{-13}$ and $10^{-14}$.

\section{Synodic and inertial frame conversions} \label{app:frame_conversion}

A state vector in the synodic rotating reference frame  \( \vec x = [x,\,y,\,\dot x,\, \dot y]^\top \) at the time $t$ can be expressed in the inertial reference frame centered at the primary body \( \vec X_1 = [X_1,\,Y_1,\,\dot X_1,\, \dot Y_1]^\top \) by applying the conversion:
\begin{equation} \label{eq:synodic_to_inertial}
    \begin{aligned}
        X_1 &= (x + \mu)\,\cos{\left(n \left(t-t_0 \right)\right)} - y \,\sin{\left(n \left(t-t_0 \right)\right)}\,,\\
        Y_1 &= (x + \mu)\,\sin{\left(n \left(t-t_0 \right)\right)} + y \,\cos{\left(n \left(t-t_0 \right)\right)}\,, \\
        \dot X_1 &= \left( \dot x - n\,y \right)\,\cos{\left(n \left(t-t_0 \right)\right)} - \left[\dot y + n\,(x+\mu) \right]\,\sin{\left(n \left(t-t_0 \right)\right)}\,, \\
        \dot Y_1 &= \left( \dot x - n\,y \right)\,\sin{\left(n \left(t-t_0 \right)\right)} + \left[\dot y + n\,(x+\mu) \right]\,\cos{\left(n \left(t-t_0 \right)\right)}\,, \\
    \end{aligned}
\end{equation}
where $n$ is the synodic frame mean motion and $t_0$ represents the initial epoch at which the two frames are aligned.

The inverse transformation from the inertial frame centered at the primary body to the synodic frame is:
\begin{equation} \label{inertial_to_synodic}
    \begin{aligned}
        x &= X_1\,\cos{\left(n \left(t-t_0 \right)\right)} + Y_1 \,\sin{\left(n \left(t-t_0 \right)\right)} - \mu\,,\\
        y &= - X_1 \,\sin{\left(n \left(t-t_0 \right)\right)} + Y_1 \,\cos{\left(n \left(t-t_0 \right)\right)}\,, \\
        \dot x &= ( \dot X_1 + n\,Y_1 )\,\cos{\left(n \left(t-t_0 \right)\right)} + (\dot Y_1 - n\,X_1 )\,\sin{\left(n \left(t-t_0 \right)\right)}\,, \\
        \dot y &= -( \dot X_1 + n\,Y_1 )\,\sin{\left(n \left(t-t_0 \right)\right)} + (\dot Y_1 - n\,X_1 )\,\cos{\left(n \left(t-t_0 \right)\right)}\,. \\
    \end{aligned}
\end{equation}

\end{appendix}
\end{document}